 \documentclass[article]{rmaa}
\usepackage{paralist}
\usepackage{natbib}
\usepackage{color}
\usepackage{graphicx}
\usepackage[referable]{threeparttablex}
\usepackage{longtable,booktabs,threeparttablex}
\usepackage{multirow, array} % para las tablas

\newcommand\vv{{\mathrm v}  }

\title{Lithium in V505 Per \\  }

\author{Gloria Koenigsberger,\altaffilmark{1}  
Werner Schmutz,\altaffilmark{2}
Catherine Pilachowski,\altaffilmark{3}
Alan R. Mej\'{\i}a-Nava,\altaffilmark{4}
Derek Sikorski,\altaffilmark{5}
Mar\'ia Cordero, \altaffilmark{6}
}

\altaffiltext{1} {Instituto de Ciencias F\'{\i}sicas, Universidad Nacional Aut\'onoma de M\'exico,
M\'exico, gloria@icf.unam.mx}
\altaffiltext{2} {Physikalisch-Meteorologisches Observatorium Davos and World Radiation Center,
               Switzerland, werner.schmutz@pmodwrc.ch}
\altaffiltext{3} {Department of Astronomy, Indiana University, Bloomington, Indiana, USA, cpilacho@iu.edu}
\altaffiltext{4} {Universidad Aut\'onoma del Estado de Morelos, Cuernavaca, Morelos,
M\'exico, alan.mejianav@icf.unam.mx}
\altaffiltext{5} {Institute for Astronomy, University of Hawai‘i, 2680 Woodlawn Drive, Honolulu, HI 96822, USA,
dsikors@hawaii.edu}
\altaffiltext{6} {Department of Astronomy, Indiana University, Bloomington, Indiana, EUA, mjcorder@gmail.com}

\shortauthor{Koenigsberger et al.}

\shorttitle{Lithium V505 Per}

\fulladdresses{

\item G. Koenigsberger, Instituto de Ciencias F\'{\i}sicas, Universidad Nacional Aut\'onoma de M\'exico,
Ave. Universidad S/N, Cuernavaca, Morelos, 62210, M\'exico, (gloria@icf.unam.mx).

\item C. Pilachowski, Indiana University, Bloomington, Indiana, EUA

\item W. Schmutz, Physikalisch-Meteorologisches Observatorium Davos and World Radiation Center,
              Dorfstrasse 33, CH-7260 Davos Dorf, Switzerland

\item A. R. Mej\'{\i}a-Nava, Universidad Aut\'onoma del Estado de Morelos, Cuernavaca, Morelos.

\item D. Sikorski, Institute for Astronomy, University of Hawai‘i, 2680 Woodlawn Drive, Honolulu, HI 96822, USA 

\item M. Cordero,  mjcorder@gmail.com

}

\listofauthors{G. Koenigsberger, C. Pilachowski, W. Schmutz, A. Mej\'{\i}a-Nava, D. Sikorski, M. Cordero}

\indexauthor{Koenigsberger, G.}
\indexauthor{Schmutz, W.}
\indexauthor{Pilachowski, C.}
\indexauthor{Mej\'{\i}a, A.}
\indexauthor{Sikorski, D.}
\indexauthor{Cordero, M.}

\resumen{
We determine the surface lithium abundance of the eclipsing binary components in V 505 Per (HIP 10961),
$A(Li)$=2.65$\pm$0.07 and 2.35$\pm$0.1, which supports the rather unexpected conclusion that their 
surface Li abundances differ. We find effective temperatures 6600 K + 6550 K  ($\sim$150 K higher than 
previously reported), which place  the stars at the hot limit of the Lithium Dip, thus aleviating the 
previously suggested discrepancy with cluster stars of similar ages and temperatures. These temperatures 
are also more consistent with the system's {\it Gaia} spectral energy distribution.  Our iron abundances,
$[Fe/H]=-0.15\pm 0.07$ and $-0.25\pm 0.1$,  agree with predictions of the higher temperatures deduced 
from our spectra and from evolutionary tracks. The rotation rate implied by our line profiles,  
12.5$\pm$1 km/s, is smaller than the synchronous rotation rate, a curious result given the circular orbit 
and the age of the system.

}      %end abstract parenthesis

\abstract{El an\'alisis de un espectro de alta dispersi\'on del sistema binario eclipsante V505 Per (=HIP 10961)
resulta en abundancias de litio consistentes con los encontrados en un estudio anterior, pero temperaturas
efectivas $\sim$150 grados mayores que en estudios previos.  Las nuevas temperaturas colocan a las dos
componentes en el l\'{\i}mite caliente del ``Lithium Dip", eliminando la inconsistencia con  resultados de
estrellas en c\'umulos de la misma edad y masa.  La abundancia superficial de litio difiere entre las dos 
componentes, sustentando resultados preliminarios previos y carece de una explicaci\'on dado que ambas
estrellas tienen casi las mismas propiedades f\'{\i}sicas. La velocidad de rotaci\'on es 12.5 $\pm$ 1 km/s,
un valor menor al valor de rotaci\'on s\'{\i}ncrona, situaci\'on curiosa dada la edad del sistema binario.
}

\addkeyword{  }
\addkeyword{Stars: binaries:eclipsing}
\addkeyword{Stars: binaries: spectroscopic}
\addkeyword{Stars: chemical abundances }

\begin{document}
\maketitle

\section{Introduction} \label{sec:intro}

Double line eclipsing binaries  are among the most important astrophysical objects
available for understanding a variety of physical processes. 
One of the many astrophysical problems they can be used to address is the one associated
with the manner in which nuclear processed material is transported from inner layers to the stellar
surface.  This is because the two stars in the system can be assumed to have formed from the
same molecular cloud and at the same time, and thus have the same age.  If they happen to have 
the same mass, they can be expected to follow identical evolutionary paths  as their interior 
mixing processes should be the same. Any difference in chemical surface abundance would serve
to quantify differences in the mixing processes.

V505 Persei (HD\,14384, HIP\,10961) is a short-period, double-line eclipsing binary system in which 
both stars are spectroscopically very similar.  The slightly brighter, larger, and more massive of 
the two is called component A, and its companion is component B.  The orbital period is well-known
to be 4.22\,d although, as noted by \citet{2021Obs...141..234S}, it was first thought
to be 2.11\,d, half the actual value, due to the nearly identical shape of the two eclipses.\footnote{Note
that  automatic period-search algorithms also commit the same mistake even in recent times \citep{2022ApJS..258...16P}}
The two components have nearly the same  mass (1.27 $M_\odot$+1.25 $M_\odot$) and radii 
(1.29 $R_\odot$+1.26 $R_\odot$) Southworth (2021, henceforth S21).
%(\citet{2021Obs...141..234S}, henceforth, S21).
The effective temperatures $T_{eff}$, determined by \citet{2008A&A...480..465T} (henceforth T08), 
are also very similar (6515$\pm$50 K and 6460$\pm$50 K). 

Whereas most of  its properties appear to be  determined to an exquisite precision, 
there is a significant spread in the published metallicity values.  T08 obtained
[M/H]=$-0.12\pm$0.03 from a $\chi^2$ model fit to the observed spectra.  Also using model fits 
to spectral observations, \citet{2013PASP..125..753B}  obtained a similar result $[Fe/H] = $-$0.15\pm$0.03.
According to T08, this metallicity is consistent with V505 Per's $\sim$1 Gyr age and the idea
that element diffusion during this time has reduced the surface metallicity from an initial
$\sim$solar  value.  \citet{2011yCat..35300138C} found $[M/H]$=$-$0.25 based on a re-assessment
of the photometric measurements listed in the \citet{2007A&A...475..519H} Geneva-Copenhagen survey 
of the Solar neighbourhood, where it was listed with $[M/H]$=$-$0.35. 
In contrast to these sub-solar values, S21 found  a slightly super-solar metal abundance 
by comparing the observed properties with theoretical mass-radius and mass-effective temperature 
diagrams.\footnote{These results implied a fractional metal abundance $Z$=0.017 and an age 1.050$\pm$0.050 Gyr. 
S21 noted that the $[M/H]\sim -0.12$ value could be excluded from his results because it corresponds to 
$Z$=0.0116 (for a heavy-element mixture such that the solar metallicity is $Z_\odot$=0.01524), a value 
outside the bounds of the range allowed by the theoretical models.}  Hence, as suggested by S21, 
a reappraisal of the the metallicity is warranted.

Another curious aspect is the conclusion reached by Baugh et al. (2013) regarding the lithium     
abundance in the two stars. They found star A to have A(Li)$\sim$2.67$\pm$0.1
and star B to have A(Li)$\sim$ 2.42 $\pm$ 0.2.  The T08 temperatures place both stars 
within the {\it Lithium Dip}, a region in the $A(Li)$ {\it vs.} $T_{eff}$ diagram within the temperature
range $\sim$6300\,K- 6600\,K, in which the surface lithium abundance is severely depleted 
compared to stars that are hotter and cooler than this temperature range \citep{1986ApJ...303..724B, 
1993ApJ...415..150T,2000A&A...357..931D}.  The Baugh et al. (2013) lithium abundance values are ~2-5 times 
larger than the detections and upper limits derived in the similar metallicity and intermediate-age open 
clusters NGC 752 and 3680, as well as the more metal-rich and younger Hyades and Praesepe.  Baugh et al. (2013) 
suggested that the higher A(Li) values supported the hypothesis that different rotational evolution of 
stars in short period binaries affects their lithium depletion, making them different from single stars.
In the case of massive stars, \citet{2023A&A...671A.139P} have concluded that the interior chemical element 
transport is not as efficient in binary star components as in their single-star counterparts in the same mass 
regime and evolutionary stage.  Thus, the question of internal mixing is relevant across the Hertzprung-Russell diagram.

In this paper we analyze a high resolution echelle spectrum obtained at orbital phase 0.97,
when the lines of both stars are well resolved, 
in order to  further constrain the  metallicity,  projected rotation velocity and
lithium abundance of both components.  In Section 2, we describe the observations and data
reduction.  In Section 3, we perform a light curve analysis of {\it TESS} data.  In Section 4, 
we analyze the $[Fe/H]$ abundances by both a curve of growth method and a detailed comparison
of theoretical spectral lines to the observed line profiles.  In Section 5 we constrain the Lithium abundance, 
in Section 6 we discuss the results and in Section 7 we summarize the conclusions.

\section{Observations}

V505 Per (=TIC 348517784) was observed by the TESS satellite (Ricker et al. 2015) in sectors 18 (2019),
58 (2022), and 85 (2024). 
The 2-minute cadence data files were downloaded from the Mikulski Archive for Space Telescopes (Swade et al. 2019) 
using search\_lightcurve (Ginsburg et al. 2019). These files contain the Simple Aperture Photometry fluxes (SAP flux) 
and the background counts as reduced by the TESS pipeline. The light curves are the result of aperture photometry, 
which gives total counts measured from the TESS images within the photometric aperture. The light curves were 
extracted and normalized using the Lightkurve Collaboration et al. (2018) software, excluding data that do not 
have a quality equal zero and epochs having background counts higher than 50000 e$^{-}$/s, where typical count 
rates of the targets are 400,000\,e$^{-}$/s. We also excluded the first part of the Sector 18
light curve earlier than BTJD 1794\,d because there the fluxes rise to 1.01 of the normalized uneclipsed level. 
It is not clear if this is a real flux change or something else, and there is no other similar deviation from a 
regularly repeating orbital variation of the light curve.

Our spectroscopic observations of V505 Per were obtained on 2011 Sept 16 and 20 at the Kitt Peak National Observatory  
Coud\'e Feed telescope using the echelle grating with a slit width of 250\,$\mu$m, Camera 5 and the F3KB CCD.  
The echelle grating provides a reciprocal dispersion of 1.9 \AA/mm at $\lambda$6697 \AA\ which, with the CCD 
0.015 mm/pixel, provides a resolution R= 110000 at this wavelength.\footnote{Willmarth, Daryl, 2.1m 
Coud\'e Spectrograph Instrument Manual, NOAO, Jan. 5, 1996.}  Flat fields and biases were obtained 
throughout the night of Sept 20, and at the beginning and end of the night on Sept 16.
%CCD has resolution of 116000.
The  Julian Dates (-2400000)  of our observations are 2455821.8953 and 2455825.9871, which
correspond to orbital phases 0.9772 and 0.9464, respectively, based on an initial epoch 
$T_0$=JD 2458798.516720 and orbital period $P$=4.2220216\,d  given by S21 (but see below).
%The T0 at the epoch of our observations was 55817.769. T0new=55821.991  (this is T0 precessed by to 2011)
The orbital phases are defined such that $\varphi$=0 corresponds to the deeper of the two eclipses, 
which occurs when star A is eclipsed by its companion.

\begin{figure} [!h !t !b]
%/Users/gloriak1/Pilacho*/Paper/input_hip10961_RVcurve
%/Users/gloriak1/Pilacho*/Spectra/Originals/input_montage_hip10961
\includegraphics[width=0.48\linewidth]{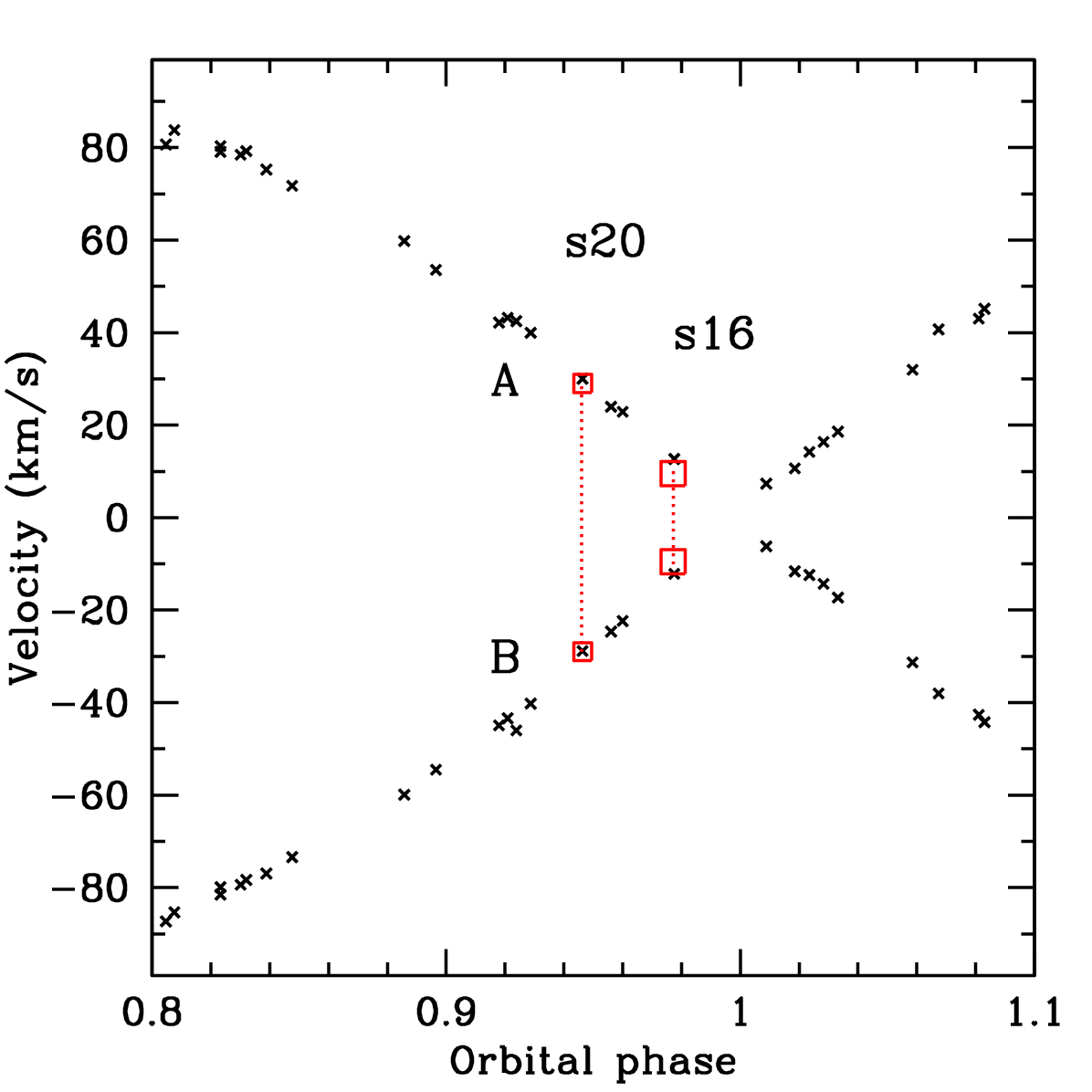}
\includegraphics[width=0.48\linewidth]{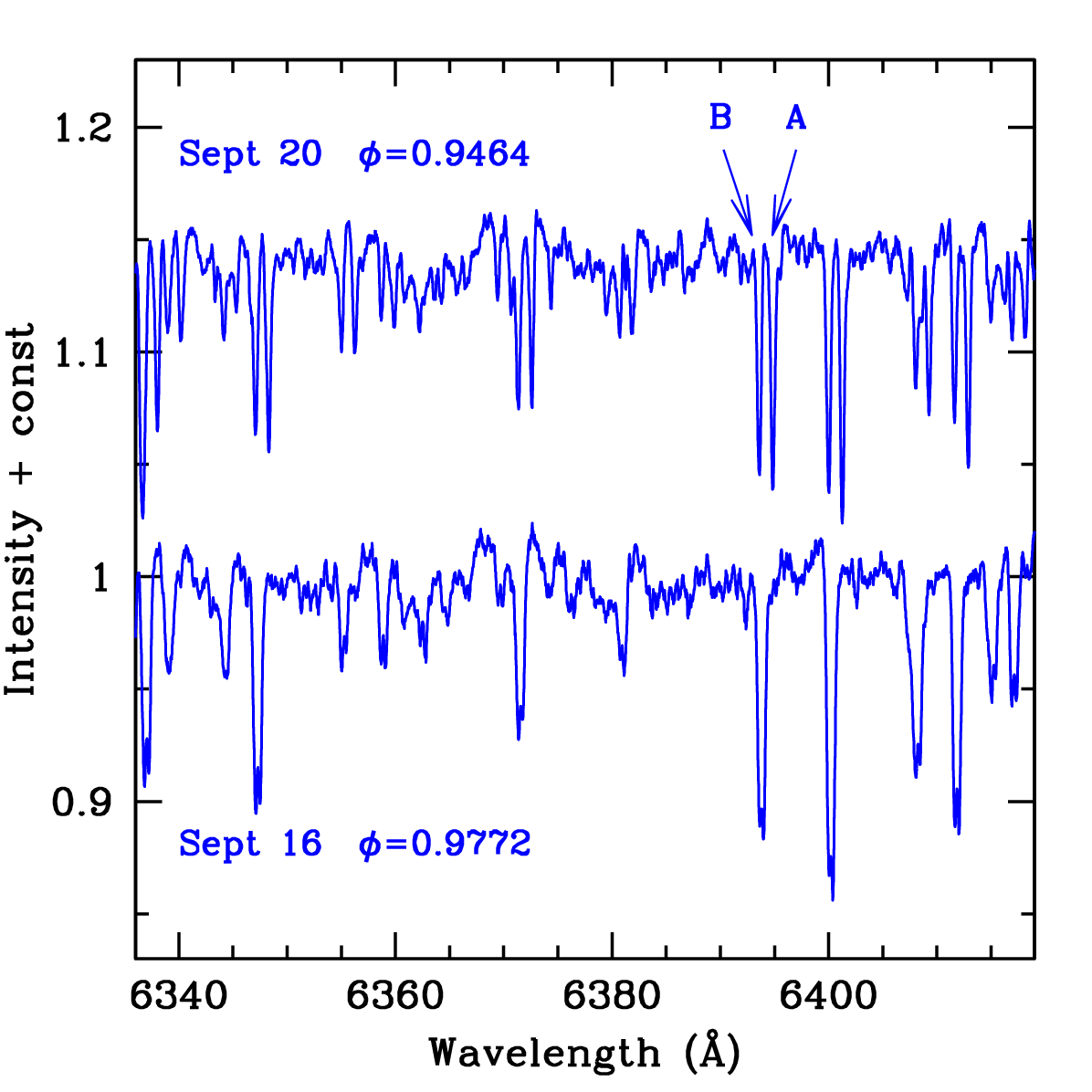}
\caption{{\bf Left:} Radial velocity curves near the $\phi$=0.00 eclipse, constructed from the data in 
the SB9 catalogue (crosses).  The dotted  vertical lines indicate the RV separation ($\Delta RV$) of
components A and B in our two spectra, which were obtained on Sept 16 and 20. At phase 0, star B is
in front of star A.  {\bf Right:} Spectra of V505 Per obtained at $\varphi$=0.9464 (20 September, top) 
and  $\varphi$=0.9778 (16 September, bottom), phases computed with Eq. 1.  A constant vertical shift in the first 
of these was introduced for clarity in the figure.  The star B wavelength scale is centered on the 
laboratory wavelength, so  star A's spectrum is shifted by $+$58 km/s with respect to that of star B on Sept. 20
(as indicated by the labels A and B). The two components are not resolved on Sept. 16.
}
\label{fig_RVcurve}
\end{figure}

The Sept 16 spectrum was obtained just before ingress of primary eclipse (first contact occurs at 
$\varphi\sim$0.985 see Fig. 2 of S21). At this phase, the lines of the two stars are barely
resolved, their centroids being separated by only 19 km/s.   The Sept. 20 spectrum was obtained sufficiently
far from conjunction for the lines of the two stellar components to be well separated (58 km/s), allowing
a more straightforward analysis by avoiding blending effects.  

The echelle spectra were reduced according to standard methods available using IRAF.  The wavelength 
calibration is not absolute, so we opted to shift the spectrum so that the absorption lines belonging
to the star that is approaching the observer are centered on the laboratory wavelengths.  In our spectra,
this is the  lower-mass component (star B), so this is the spectrum that is centered on the laboratory
wavelengths. The primary's lines appear shifted by $+$58 km/s.

A compendium of  V505 Per radial velocities is available in The Ninth Catalogue of Spectroscopic 
Binary Orbits \citep{2004A&A...424..727P}\footnote{http://sb9.astro.ulb.ac.be}, from which we plot 
in Fig.~\ref{fig_RVcurve} the data for the orbital phase interval 0.8 -- 1.1.  The
location corresponding to our two observations is indicated.  Since we only have relative RV measurements,
we connect the data points for the primary and the secondary with a dotted line and shift them vertically
on this plot to show that the relative RVs are fully consistent with the published radial velocity curves.

The standard method to correct for the echelle blaze was applied in order to normalize each of the
spectral orders.  Corrections for a high-frequency intensity oscillation along the normalized orders
were corrected manually by fitting a high-order polynomial to the visually fitted continuum level.  This
resulted in spectral orders where the continuum generally lies at an intensity level of unity $\pm$1\% $-$ 2\%.
A sample of the spectra is shown in Fig.~\ref{fig_RVcurve}. The echelle spectra have S/N$\sim$100 
per 0.02 \AA\ pixel.  For the analysis, we applied a boxcar 5-point smoothing to enhance the S/N.

\section{Light curve analysis}

The TESS light curves comprise 15 primary minima and 14 secondary minima.  The primary minima have
a depth of $\sim$61.8\,\% of the out of eclipse light level, and the 14 secondary minima have a depth 
of $\sim$62.8\,\%. The individual light curves show a scatter in the depths of $\sim$0.1\,\%.

The epochs of light curve minima were measured with the method proposed by \citet[]{Kwee1956}.
The average estimated uncertainty of the method is 0.8\,sec, and standard deviation of the observed minima 
from a fitted straight line to the epochs of minima timings is 1.1 \,sec.

The fits to both primary and secondary minima yield the same period within the uncertainties, from which the
following ephemeris is derived:

\begin{equation}
\label{Eq1}
\mathrm{BJD\,(Min\,I)} =  2458798.51672 (\pm 3\,10^{-5})  + 4.22201933 (\pm 2\,10^{-8}) \times E
\end{equation}

\noindent where $E$ is an integer number corresponding to the orbital cycle.
The zero epoch agrees with the time determined by S21 but the uncertainty of our determination is a factor six larger. The period of our determination is a factor hundred more precise than the one given by S21 because of the larger time base available to us.

We analyzed the folded light curve from all three TESS sectors. In a partially eclipsing systems,
only the sum of the stellar radii of the two stars can be derived from the light curve, but not their ratio.
Hence,  we first explored the possible range of radii ratios that could reproduce the observed light
curve and found that radii ratios  in the $rr=R_B/R_A$ in the range 0.85 to 1.0 yield identical light curves, all of which
fit  the observed light curve accurately.  Such a broad range in $rr$ values introduces a large uncertainty.
Thus, in order to further constrain the possible values, we analyzed the properties of evolutionary tracks.

\begin{figure} [!h]
%Figure made by Werner.                                  
\includegraphics[width=0.78\linewidth]{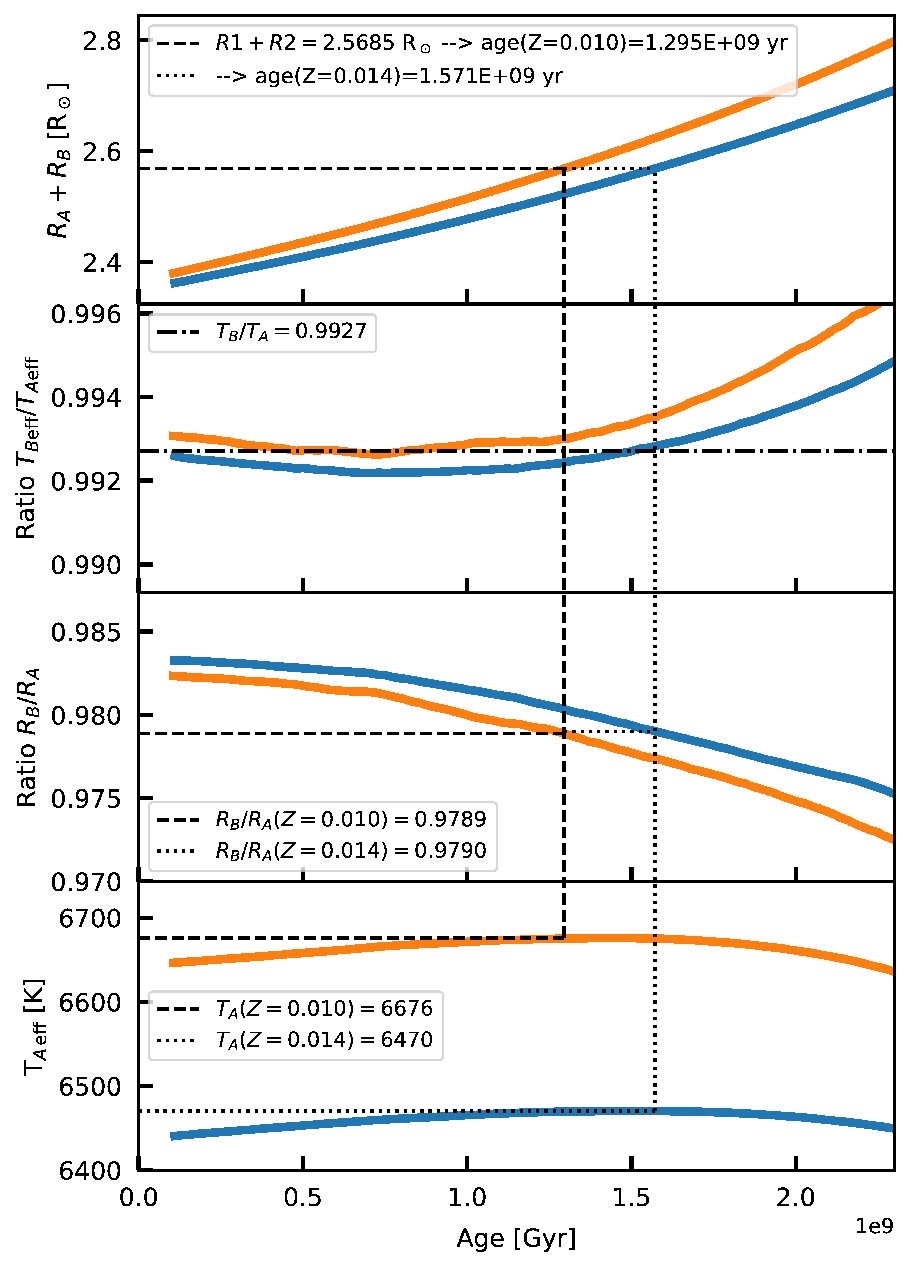}
\caption{Evolutionary tracks for a stellar pair of $M_1=1.2747 + M_2=1.2579$ $M_\odot$ 
with the solar metallicity (Z=0.014, blue) and for Z=0.010 (orange). The abscissa is the age starting with the ZAMS.
The ordinate is as follows from the top to the bottom panel: the sum of the two stars' radii, $R_A +R_B$; 
the effective temperature ratio, $T_{Beff}/T_{Aeff}$; radii-ratio, $R_B/R_A$; and the value of
$T_{Aeff}$. The $R_A +R_B$ value is obtained directly from the solutions to the light curve and radial velocity curve,
and its value for V505 Per then determines the age which, in turn, constrains
the remaining parameters.  The dash line gives their value for $Z$=0.010 and the dotted line 
for $Z$=0.014 (close to solar, $[Fe/H=-$0.037).
% The relation between z and [Fe/H] is given in C. del Burgo & Allende-Prieto (2018, MNRAS, "Testing models....PARSEC"):
%const=0.01524                  #for their solar mix 
%FeOverH = np.array(0.00,-0.0368, -0.1, -0.146, -0.2, -0.25])          #for the metallicities I used in the paper
%result = const * 10**FeOverH     =[0.01524   0.01400184 0.01210556 0.01088892 0.00961579 0.00857008]
}
\label{fig_evolution_tracks}
\end{figure}

We obtained the evolutionary tracks  for  1.2745 $M_\odot$ (star A) and 1.2577 $M_\odot$ (star B) by 
interpolating the models given by \citet{2012A&A...541A..41M} for solar 
metallicity ($Z$=0.014) and metallicity corresponding to $[Fe/H]=$-0.146 ($Z$=0.010).   We extracted
from these models the radii and $T_{eff}$ as a function of age, from which we then calculated $R_B/R_A$,  
$R_A+R_B$ and $T_{eff}$ also as a function of age.  The result is shown in Fig.~\ref{fig_evolution_tracks}.

\begin{center}
\begin{longtable}{llll}
\caption{Parameters from {\it TESS} light curve fit and literature  \label{tab_werner}} \\
\toprule
% The data in this table come Werner's light curve analysis 
Parameter     & Value          &  Notes      \\
\hline
$q$                    & 0.9868 (fixed) & RV curve S21 \\
$a\sin(i)$ ($R_\odot$) & 14.974 (fixed) & RV curve S21 \\
$i$ deg                &87.859 (fitted) & light curve \\
$a$  ($R_\odot$)       & 14.984         & derived from above \\
($R_A+R_B$)/$a$        & 0.1714 (fitted)& light curve  \\
$R_A+R_B$  ($R_\odot$) & 2.5685         & derived from above  \\
$T_B/T_A$              & 0.9928 (fitted)& light curve  \\
$R_B/R_A$              & 0.97895 (free) & Ev. Tracks   \\
\midrule
\bottomrule
\end{longtable}
 \begin{tablenotes}
  \item{} {           }
 \end{tablenotes}
\end{center}

From the light curve solutions we get the value of the sum of the stellar radii in units of the
projected orbital separation, $r_A+r_B$, where $r_i=R_i/a$, with $a$ the semi-major axis of the orbit,
which is obtained from the solution of the radial velocity curve  ($a$=14.984 $R_\odot$, S21's Table V).
%S21's light curve solution yielded the sum of fractional radii $(R_A+R_B)/a)=0.170906$ and, thus, $R_A+R_B=$2.5609.  
Our light curve solution yields $(R_A+R_B)/a$=0.1714.  With the above value of $a$, this yields $R_A+R_B=$ 2.5685$\pm$0.001,
where the uncertainty corresponds to the different results obtained from fitting the mean of all {\it TESS} 
sectors.\footnote{Results for the individual sectors are as follows: 
($T_B/T_A$, $R_A+R_B$, $i$, $M_A+M_B$)=(0.99270, 2.5691, 87.869deg, 2.53) for Sector 18; =(0.99272, 2.5684, 87.851, 2.53) for Sector 58; 
=(0.99280, 2.5694, 87.851, 2.53) for Sector 85. All fits performed with PHOEBE used $rr$=0.979}
The top panel of Fig.~\ref{fig_evolution_tracks} shows that $R_A+R_B=$ 2.5685 corresponds to an age 1.295 Gyr if the
stars have $Z$=0.010 and 1.571 Gyr if they have approximate solar metallicity.  Inspection of the second and third 
panels in this figure shows that the predicted temperature ratio and radius ratio do not have a very strong
dependence on metallicity. For both metallicities $T_{Beff}/T_{Aeff}=$0.9927, similar to the 
0.9328 value obtained from the light curve fit, and $rr$=0.97895$\pm$0.00005. 

Whereas the temperature ratio is not very sensitive to metallicity, the actual temperature has a 
very strong dependence on metallicity. The bottom panel of Fig.~\ref{fig_evolution_tracks} illustrates 
star A's effective temperature as a function of age for the two analyzed metallicites. For an approximately
solar metallicity, the temperatures are within the range that was determined by T08 (with the S21 uncertainties) 
but the system is significantly older (1.57 Gyr instead of $\sim$1 Gyr).  However, if instead the $[Fe/H]$ values are as low as 
determined by Baugh et al (2013) and Casagrande et al., then  $T_{eff}$ must be higher than determined by T08.  
The bottom panel of Fig.~\ref{fig_evolution_tracks} indicates that $T_{Aeff}\sim$6676\,K which, from the temperature
ratio means that $T_{Beff}\sim$6627\,K.
 
%%%%%%%%%%%%%%%%%%%%%%%%%%%%%%%%
\begin{center}
\begin{longtable}{llll}
\caption{Fundamental parameters of V505 Per\label{tab_southworth}} \\
\toprule
% The data in this table come S21s (2021) Tables 
Parameter           &Literature         & \multicolumn{2}{c}{This paper}      \\
\hline
$P_{orb}$ (d)      & 4.2220216(.0000023)        &4.22201933(10$^{-8}$)&\nodata       \\
$M_A$ ($M_\odot^N$)& 1.2745(.0036)$^{(a)}$      &\nodata             & \nodata       \\
$M_B$ ($M_\odot^N$)& 1.2577(.0030)$^{(a)}$      &\nodata             & \nodata      \\
$rr$               & 0.9788(.0019)$^{(b)}$      & 0.9789             & 0.9790      \\
light ratio        & 0.9367(.0037)$^{(c)}$      & \nodata            &\nodata     \\
$i$                & 87.9166(.0030)$^{(c)}$     & 87.859             &\nodata      \\
$R_A$ ($R_\odot^N$)& 1.2941(.0016)$^{(d)}$      & 1.2982             &\nodata       \\     
$R_B$ ($R_\odot^N$)& 1.2637(.0017)$^{(d)}$      & 1.2702             &\nodata        \\   
$\log(g)_A$($\log[cgs]$)& 4.3194(.0010)$^{(e)}$ & \nodata            &\nodata       \\      
$\log(g)_B$($\log[cgs]$)& 4.3343(.0010)$^{(e)}$ & \nodata            &\nodata      \\
$[Fe/H]_A$         & -0.25 $^{(f1)}$            & Tab. 3             & Tab. 3      \\
                   & -0.12(.03)$^{(f2)}$        & \nodata            &               \\
                   & -0.15(.05)$^{(f3)}$        & \nodata            &               \\
$[Fe/H]_B$         & -0.12(.03)$^{(f4)}$        & Tab. 3             & Tab. 3      \\
$z$                &  0.017$^{(f5)}$            & 0.014              &0.010              \\
$T_{Aeff}$(K)     & 6512(50)$^{(g)}$            & 6470               & 6676                 \\
$T_{Bff}$(K)     &  6460(50)$^{(g)}$            & 6423               & 6627                  \\
$V\sin(i)_A$ (km/s)  &  15.3(1.0)$^{(h)}$       & 12.5 (1)           & 12.5(1)               \\
$V\sin(i)_B$ (km/s)  &  15.4(1.0)$^{(h)}$       & 12.5 (1)           & 12.5(1)               \\
$\xi_{th}$ (km/s)   & 1.7$^{(j)}$               & \nodata            & \nodata               \\
$log(L_A/L_\odot^N)$& 0.434(.0013)$^{(k)}$      & 0.425              & 0.479                \\
$log(L_B/L_\odot^N)$& 0.399(.0013)$^{(k)}$      & 0.393              & 0.448                \\
Age (Gyr)           &  1.050(0.050)$^{(f5)}$    & 1.571              & 1.295            \\
$D (pc)$            & 61.19 (.62)$^{(m)}$         &                    &                     \\ 
$D (pc)$            & 62.03 (.10)$^{(n)}$         &                    &                     \\
\midrule
\bottomrule
\end{longtable}
 \begin{tablenotes}
  \item{} {(aa) $M_\odot^N$, $R_\odot^N$, $L_\odot^N$ are the nominal solar units given by IAU 2015 Resolution
B3 \citep{2016AJ....152...41P}}  % (A. Pr\~sa et al., AJ, 152, 41, 2016)
  \item{} {(a) Masses from the radial velocity curves solution and orbital inclination=87.916 deg.}
  \item{} {(b) Radii ratio from the light curve fit.}
  \item{} {(c) Light ratio from the light curve fit.}
  \item{} {(d) Stellar radii from the light curve and radial velocity solutions. T08 values are
the same within their uncertainties.}
  \item{} {(e) The S21 et al. values derive from the masses and radii. The T08 values
are from the spectral fits. T08 values are the same within their uncertainties.}
  \item{} {(f1) \citet{2011yCat..35300138C}}
  \item{} {(f2) T08  from their spectrum at orbital phase 0.497, when star A nearly totally 
eclipses its companion. }
  \item{} {(f3) Baugh et al. (2013) }
  \item{} {(f4) T08              }
  \item{} {(f5) S21 isochrone fitting; z=0.017 corresponds to $[Fe/H]=+$0.05, using $z=0.01524\times10^{[Fe/H]}$,
assuming a solar-proportional mix, see \citet{2018MNRAS.479.1953D}. }  %This assumes [Fe/H]~[M/H], and $[\alpha/Fe]$=0.
  \item{} {(g) Effective temperatures from T08  based on Kurucz model spectra fits to the data, 
with uncertainty values as given by S21. }
  \item{} {(h1) Projected equatorial rotation velocity from T08 for star A based on their spectrum at orbital
phase 0.497.}
 \item{} {(h2) Projected equatorial rotation velocity from T08 for star B based on an iterative orbital
solution for which the difference in the temperature between the primary and the secondary stars and fraction of
the combined system light due to the two components was calculated.}
  \item{} {(j) Microturbulent speed as derived by Baugh et al. (2013)}
  \item{} {(k) Luminosity from the deduced $T_{eff}$ and radius values. S21 notes that the distance
implied by these luminosities ``is slightly shorter than that obtained from the Gaia EDR3 parallax, a 
discrepancy most easily explained by uncertainty in the 2MASS K-band apparent magnitude."}
\item{} {(m) S21}
\item{} {(n) {\it Gaia} (re-interpreted) EDR3 parallaxes, C. A. L. Bailer-Jones et al., AJ, 161, 147, 2021. }
 \end{tablenotes}
\end{center}

\section{Iron abundance}

The chemical abundance analysis of a binary system requires knowledge of the effective temperature
($T_{eff}$), logarithm of the surface gravity ($\log(g)$), each star's contribution to the
continuum spectral energy distribution ($w_A$ and $w_B$), and the  microturbulent velocity ($\xi_{th}$).
The methods to determine the chemical abundances rely heavily on the use of theoretical
stellar atmosphere models.  The first is the curve-of-growth method. It uses the equivalent widths of
numerous absorption lines measured on the observed spectrum which are compared to those of a grid of
stellar atmosphere models.  The second is a direct comparison of the observed absorption line spectrum
to synthetic spectra predicted by the grid of theoretical stellar atmosphere models. Both methods require
precise values of the relative continuum contributions from star A and star B,  $w_A$ and $w_B$.

Each star's contribution to the normalized continuum is obtained from the $T_{eff}$ and $rr$ values as
follows. Defining a weight $w_i$=$L_i/(L_A+L_B$), with $L_i$ the luminosity of each star (i=A, B), and assuming
that the black-body function is a valid approximation for the visual portion of the spectral energy
distribution, we can write $w_A/w_B$=$(R_A/R_B)^2(T_A/T_B)^4$, with $w_A+w_B$=1.

The results shown in Fig.~\ref{fig_evolution_tracks} indicate that both $rr$ and $T_A/T_B$ are insensitive
to metallicity.  Using their values as listed in Table~\ref{tab_werner}, we find $w_A/w_B$=1.0740, from
where $w_A$=0.518 and $w_B$=0.482.  These values are used for the analyses that follow.

\subsection{Equivalent widths of Fe lines}

We measured the equivalent widths ($W_\lambda$) for 26 \ion{Fe}{1} lines in the 
wavelength range $\lambda\lambda$6042-6718  in the 20 September spectrum. For comparison, we chose
stellar atmosphere models having the T08 and S21 effective temperature, surface gravity 
and relative luminosities, and the Baugh et al. (2013) microturbulent speed (see Table~\ref{tab_southworth}). 
The \ion{Fe}{1} abundance was determined with the {\it abfind} driver in  the LTE spectral synthesis 
and line analysis code {\it MOOG} (Sneden  2012).\footnote{http://www.as.utexas.edu/~chris/moog.html}  
{\it MOOG} uses a carefully curated list of lines of well-determined $gf$ values to synthesize a spectrum 
from a stellar atmosphere model.  The equivalent-width method for Fe-abundance determinations has the 
advantage that it does not depend on the rotation velocity. 

We tested models with [M/H] at $-$0.1, $-$0.2, $-$0.3, and $-$0.4 dex.  This resulted in abundances 
7.286, 7.285, 7.284, 7.282, respectively for star A and 7.212, 7.212. 7.211, 7.209 for star B.  Thus, 
for each star we get an average $\pm$s.d. of $\log(Fe)_A$=7.284$\pm$0.002 and $\log(Fe)_B$=7.211$\pm$0.001.

We also ran {\it MOOG} with the the solar equivalent widths for the same list of lines, and a solar model
at 5780 K and log g = 4.4 which was computed with the same program interpolating in the MARCS model
grid. Many of the lines are rather strong in the Sun, and give a lower abundance than the weaker lines.
We eliminated lines stronger than Log $W_\lambda$ = $-$4.8 ($\sim$100 mA), and got an abundance of
$\log(Fe)_\odot$ = 7.43 from this set, using a microturbulence of 1 km/sec (which is the accepted value
for center-of-disk, consistent with the McMath atlas \citep{1998assp.book.....W}).
%(Wallace, L., Hinkle, K., and Livingston, W.C. 1998, National Solar Observatory Technical Report 98-
%001, An Atlas of the Spectrum of the Solar Photosphere from 13,500 to 28,000cm-1 (3570 to 7405 A)).
%Assuming log(H)=log(H_\odot)

With  the definition $[Fe/H]=\log(Fe/H)-\log(Fe/H)_\odot$ and assuming the same hydrogen abundances 
$\log(H)=\log(H)_\odot$, the above values of $\log(Fe)$ correspond to $[Fe/H]= $-$0.146$ for star A 
and $[Fe/H]= $-$0.219$ for star B.  We adopt an uncertainty of $\pm$0.07 dex due primarily to the uncertainty 
in the continuum placement.

Our star A iron abundance is in excellent agreement with that of Baugh et al. ($-$0.15$\pm$0.03)
and T08 ($-$0.12$\pm-$0.03), both of which are based on comparisons to \citet{1992IAUS..149..225K} model
atmosphere spectra. The iron abundance that we derive for star B is marginally lower,
and more in line with the $-$0.25 value reported in \citet{2011A&A...530A.138C}.  It is even closer to the
\citet{2011A&A...530A.138C} value if a solar microturbulence of 0.8 km/sec is assumed (which gives a flatter 
dependence on line strength and a solar abundance of 7.47) resulting in  $[Fe/H]=-$0.186 and $-$0.259
for star A and B, respectively.  Column 3 of Tab.~\ref{table_best_fit} summarizes the results
of this experiment.

\subsection{Modeling absorption-line profiles}

Synthetic spectra in the $\lambda\lambda$ 6388-6418 \AA\ region were generated
using the {\it Pymoogi} Python wrapper for {\it MOOG}.\citep{2017AAS...23021607A}.\footnote{github.com/madamow/pymoogi}
We used  MARCS stellar atmospheres models \citep{2008A&A...486..951G}, interpolating the available grid
when needed, to produce synthetic spectra.  Each model is characterized by the effective temperature
$T_{eff}$, logarithm of the surface gravity $\log(g)$, and heavy element abundance $[Fe/H]$.  Our grids 
of models were generated for $T_{eff}$ in the range 6450\,K to 6800\,K, and $[Fe/H]$ in the range $-$0.4 to $-$0.1, with
fixed  $\log(g)$, and $\xi_{th}$ as listed in Column 2 of Table~\ref{tab_southworth}.

\begin{figure} [!h]
%/Users/gloriak1/Pilachowski/Spectra/input_montage_Fe_abundance, Spectra/input_montage_Fe_abundance_ratios
\includegraphics[width=0.48\linewidth]{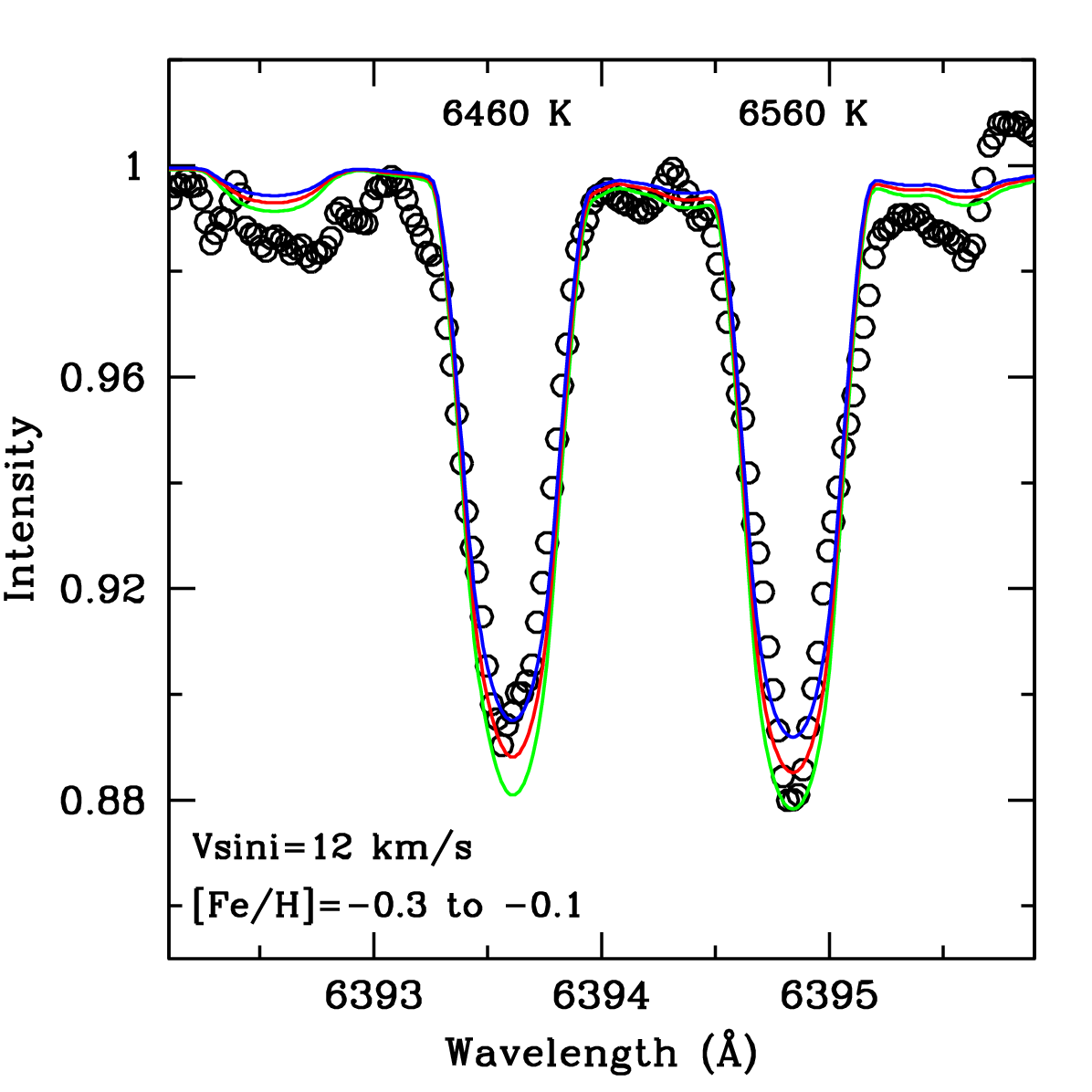}
\includegraphics[width=0.48\linewidth]{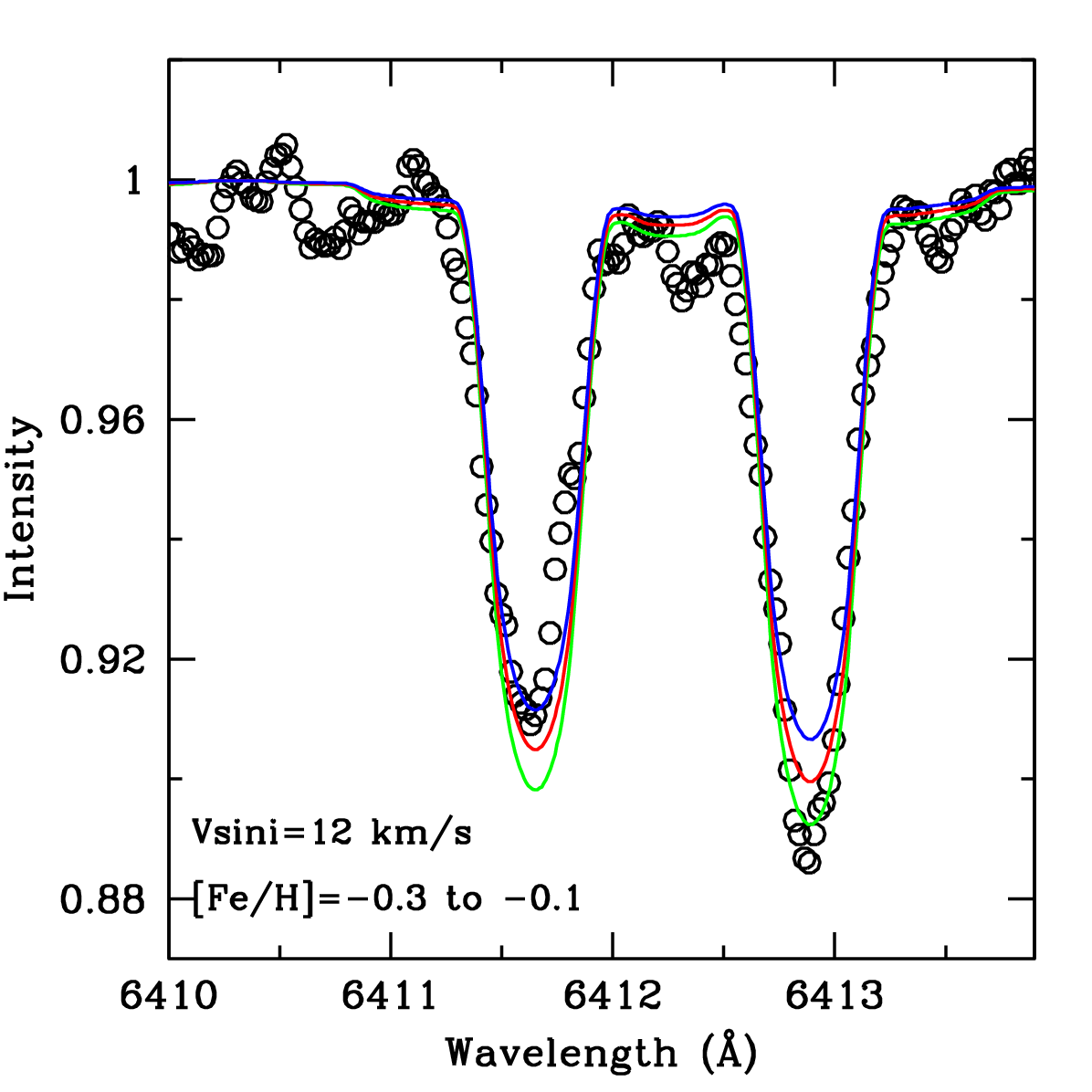}
\includegraphics[width=0.48\linewidth]{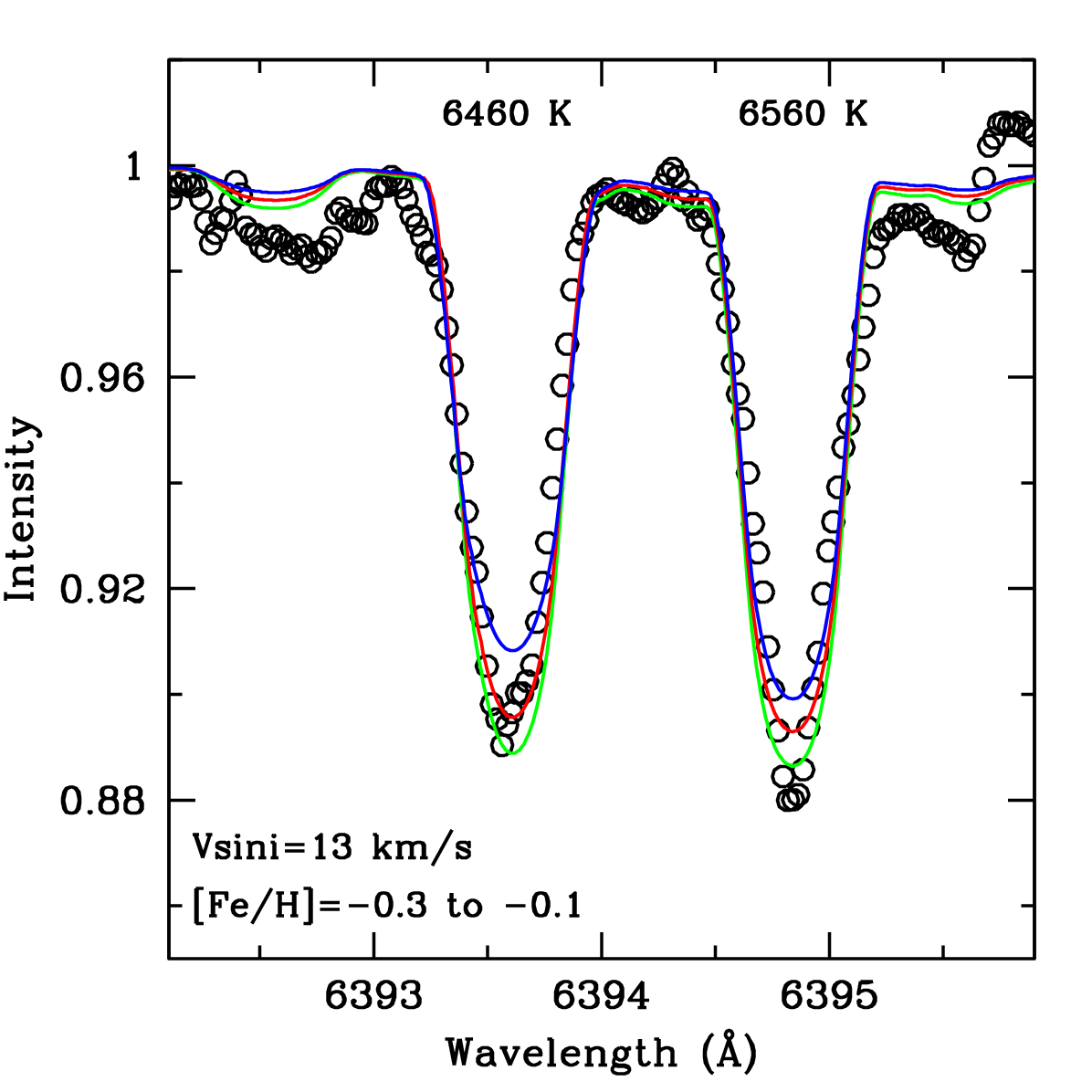}
\includegraphics[width=0.48\linewidth]{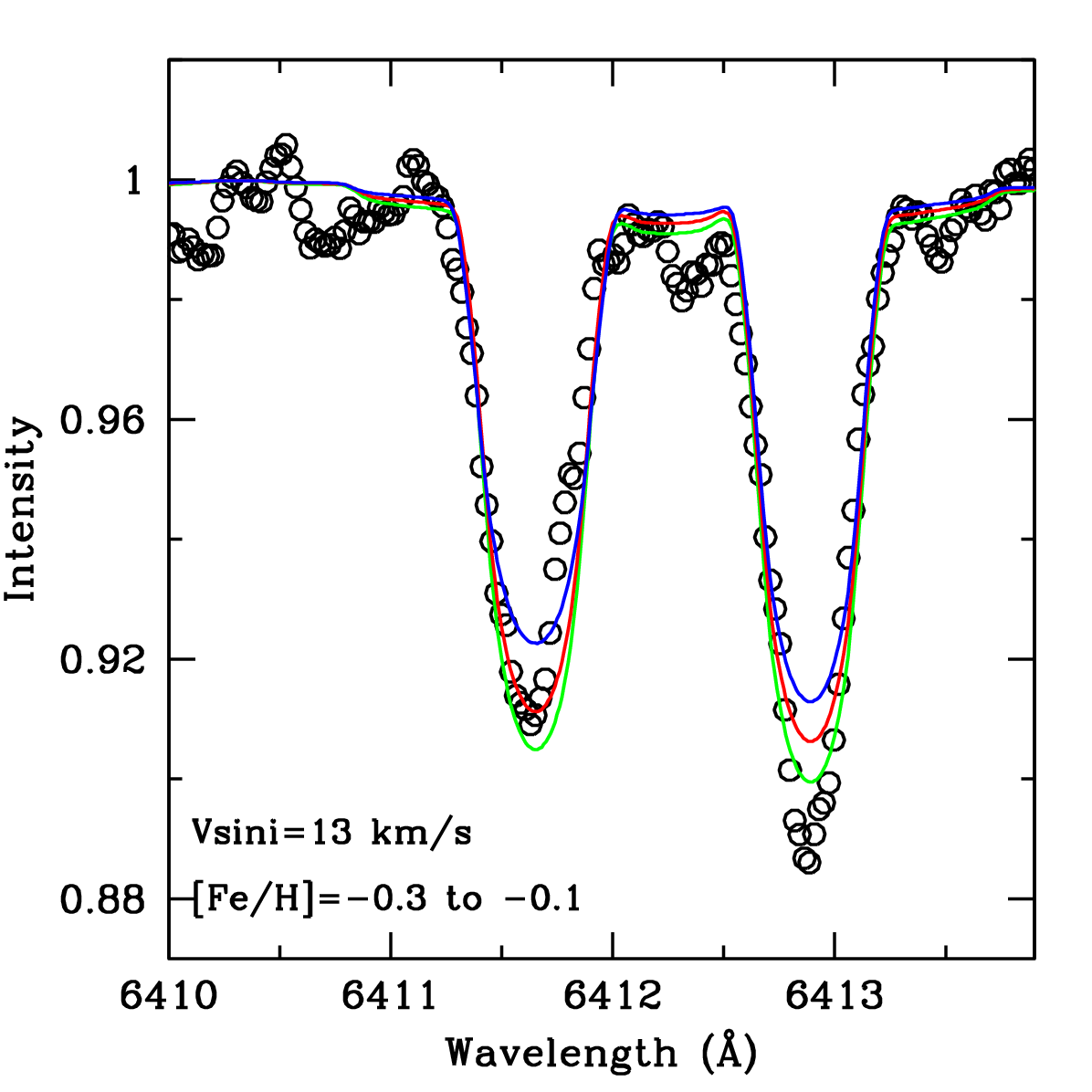}
\caption{Determination of $\vv\sin(i)$:  Shown are the rotationally-broadened synthetic line profiles from 
model atmospheres with $T_{eff}$, $\log(g)$, and $\xi$ values from column 2 of Table~\ref{tab_southworth},
and for $[Fe/H]= -0.1$ (green), $-0.2$ (red) to $-0.3$ (blue).  Rotational broadening $\vv\sin(i)$=12 km/s (top) 
and 13 km/s (bottom) compare most favorable to the 5-point smoothed observations (dots).  
}
\label{fig_testing_Vsini}
\end{figure}

The single-star spectra produced by MOOG were scaled and combined to synthesize V505 Per's double-line
spectrum using the scaling factors $w_A$=0.518, $w_B$=0.482 as discussed above.

A detailed comparison between synthetic and observed spectra requires that rotational broadening be applied to the
synthetic spectral lines.  In order to constrain the value of $\vv\sin(i)$ we used the {\it rotBroad} function 
included in the {\it pyAstronomy} package\footnote{pyastronomy.readthedocs.io/en/latest/pyasIDoc/asIDoc/rotBroad.html,
github.com/sczesla/PyAstronomy/blob/master/src/pyasl/asl/rotBroad.py} to broaden a composite
synthetic spectra to projected rotation velocities in the range 10 km/s to 14 km/s. A linear limb-darkening
coefficient 0.26 was adopted from S21.  Fig.~\ref{fig_testing_Vsini}  compares  the observed 
$\lambda$6393.6 and $\lambda$6411.6 lines to the synthetic spectra broadened to $\vv\sin(i)$=12 km/s and 13 km/s.
These two spectral lines are relatively free from blending effects. We find that the higher speed yields a 
slightly better fit to the profiles of star A, while the lower speed is slightly better for star B, but both
speeds are viable, within the uncertainties of the data. Hence, we adopt $\vv\sin(i)$=12.5 $\pm$1 km/s as the
most likely  rotation speed of both stars.  It is noteworthy that this value is smaller than the 15 km/s synchronous 
rotation velocity which was determined by T08, a discrepancy that we attribute to the lower spectral resolution
of the T08 data compared to our echelle spectra.

For $\vv\sin(i)$ between 12 and 13 km/s,  Fig.~\ref{fig_testing_Vsini} shows that the line profiles are 
consistent with $[Fe/H]=-$0.10$\pm$0.05 (star A) and $-$0.25$\pm$0.1 (star B), similar to the result 
obtained from the equivalent widths.  

\begin{figure} [!h]
%/Users/gloriak1/Pilachowski/Spectra/input_
\includegraphics[width=0.48\linewidth]{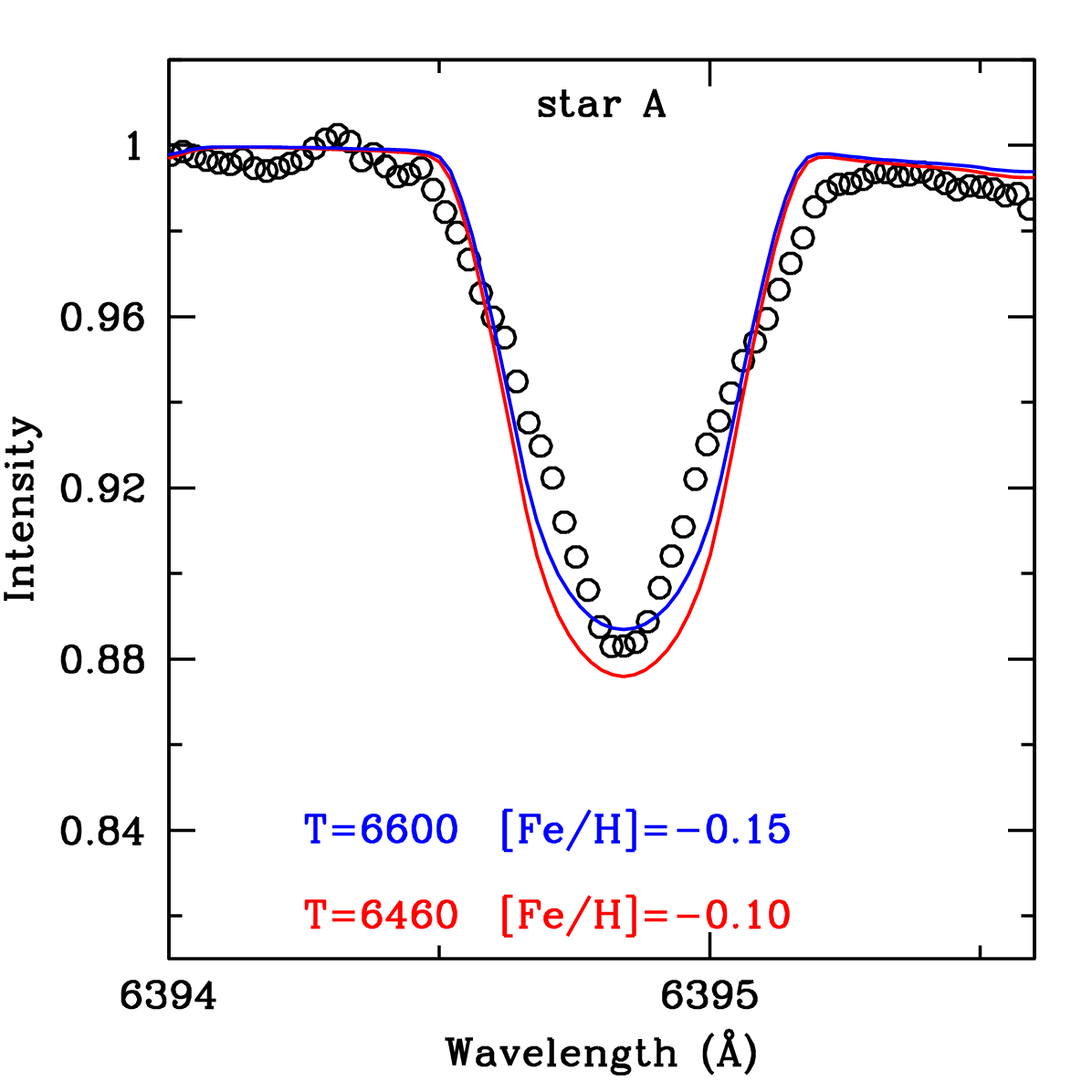}
\includegraphics[width=0.48\linewidth]{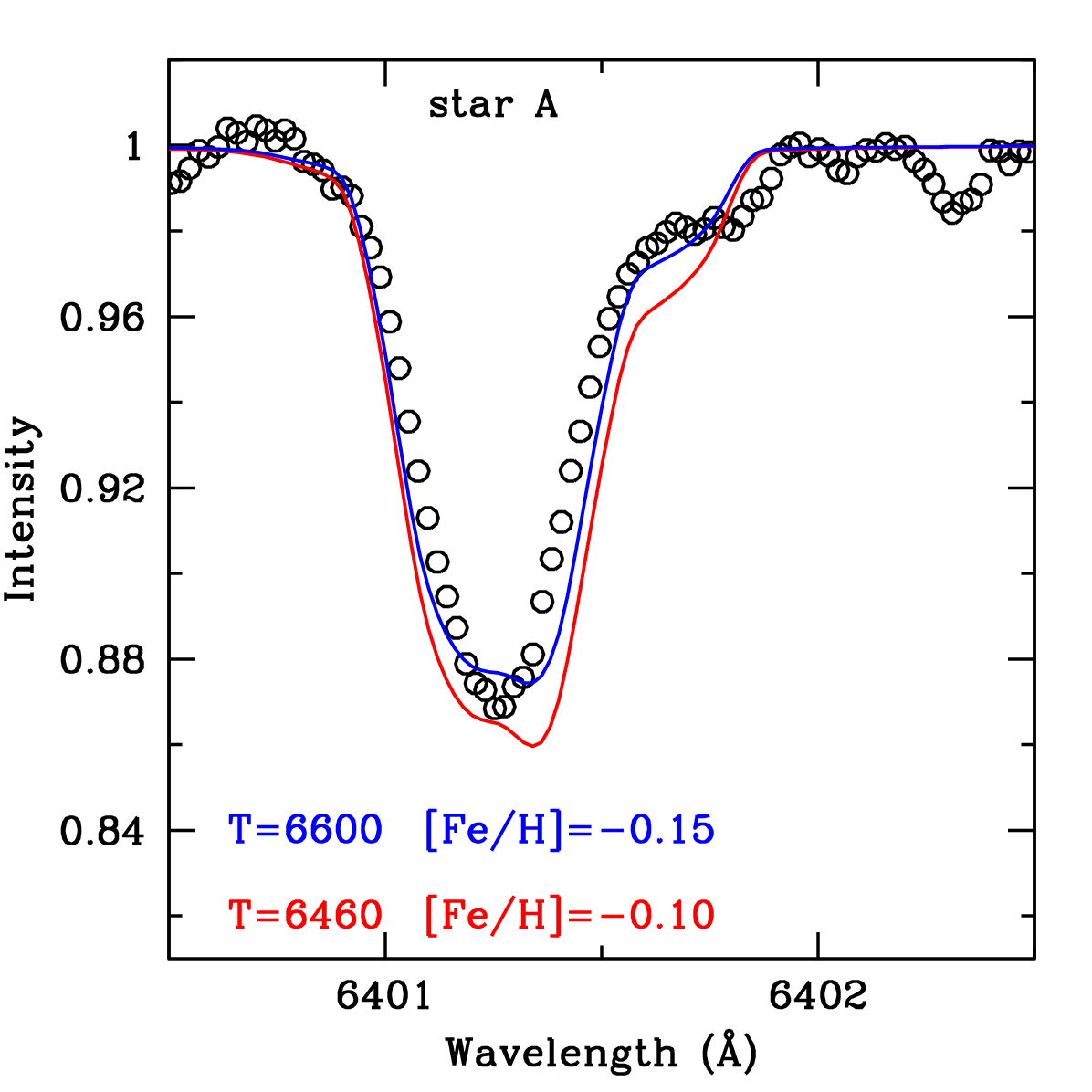}
\includegraphics[width=0.48\linewidth]{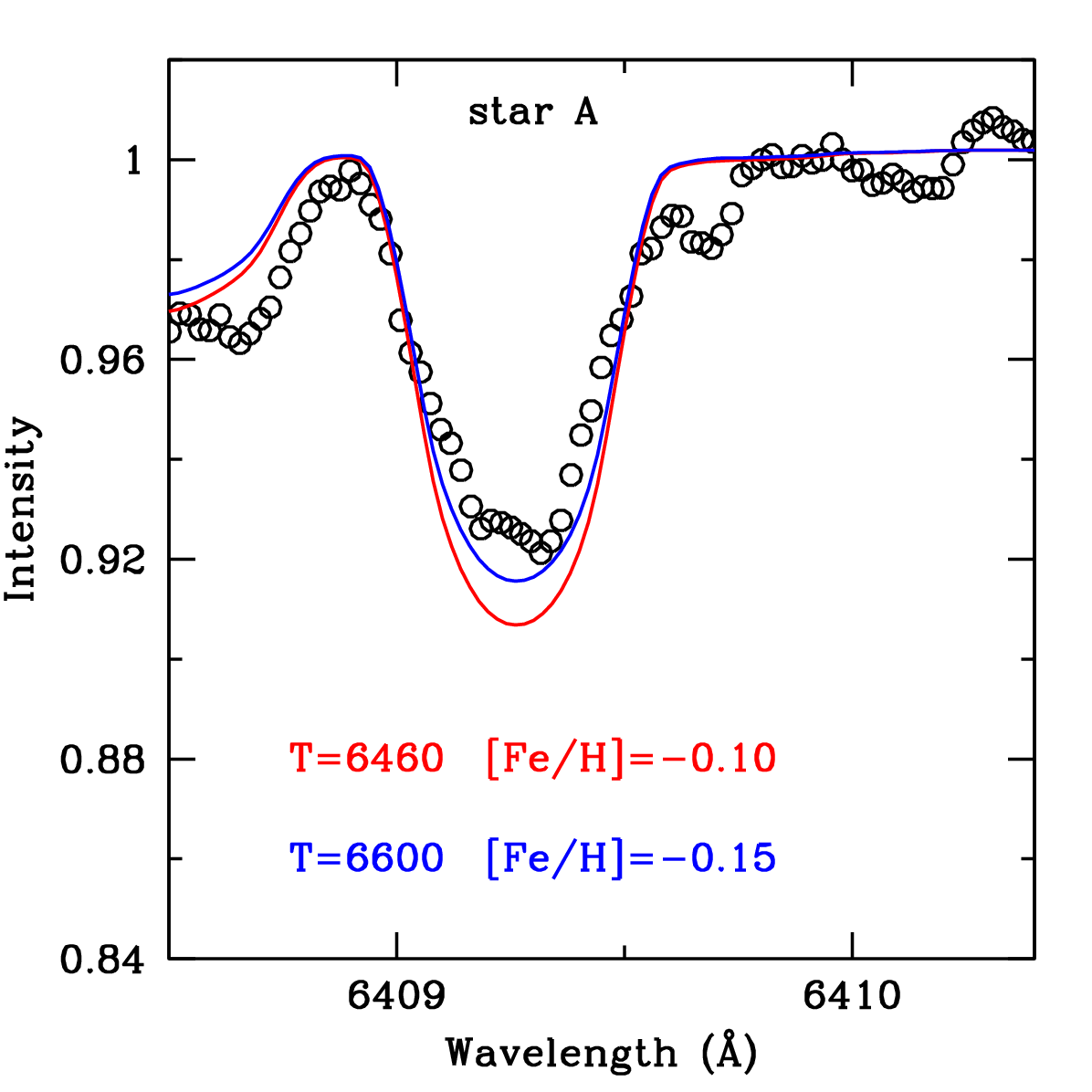}
\includegraphics[width=0.48\linewidth]{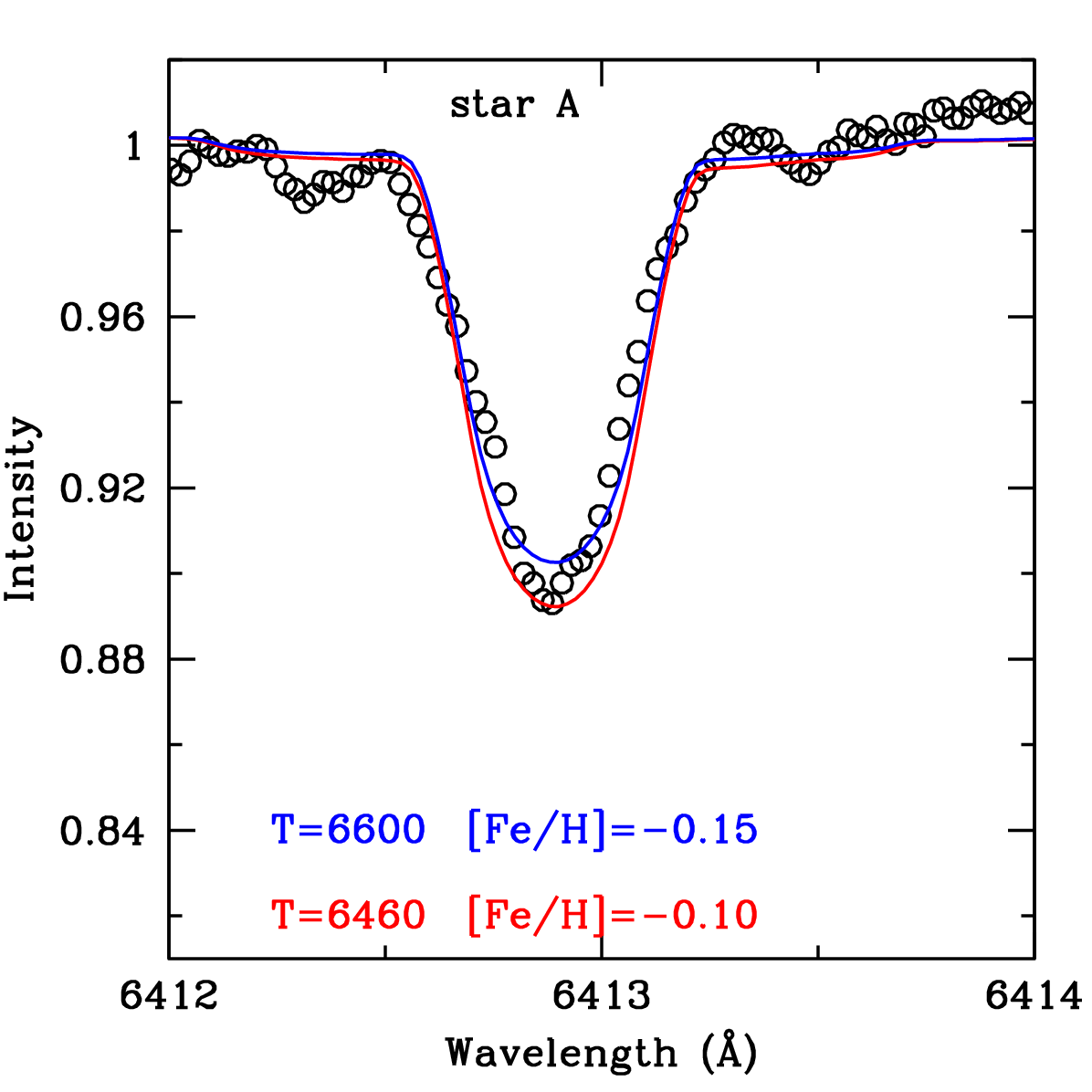}
\caption{Line line profiles of the primary star (dots) compared to model spectra with ($T_{eff}$, $[Fe/H]$)=(6460 K, $-$0.10) (red)
and ($T_{eff}$, $[Fe/H]$)=(6600 K, $-$0.15) (blue).  The model line profiles are broadened to $\vv\sin(i)$=12 km/s.   The $\lambda$6413
line is rather insensitive to $T_{eff}$, while $\lambda$6400.3 shows a relatively strong temperature sensitivity.
This figure shows that models with $[Fe/H]>-$0.1 would not adequately represent the observations. 
}
\label{fig_compare_Teff_starA}
\end{figure}

\begin{figure} [!h]
%/Users/gloriak1/Pilachowski/Spectra/input_
\includegraphics[width=0.48\linewidth]{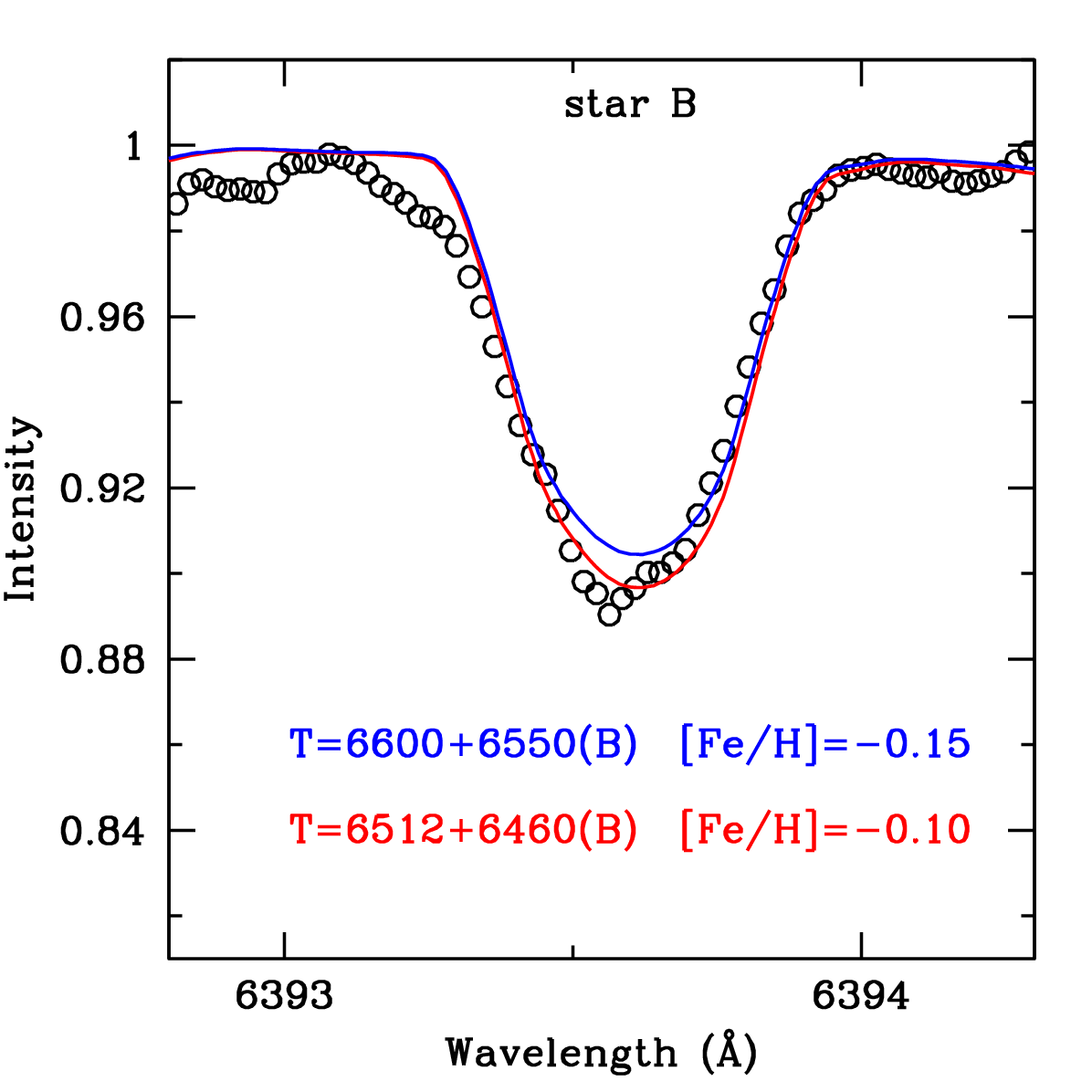}
\includegraphics[width=0.48\linewidth]{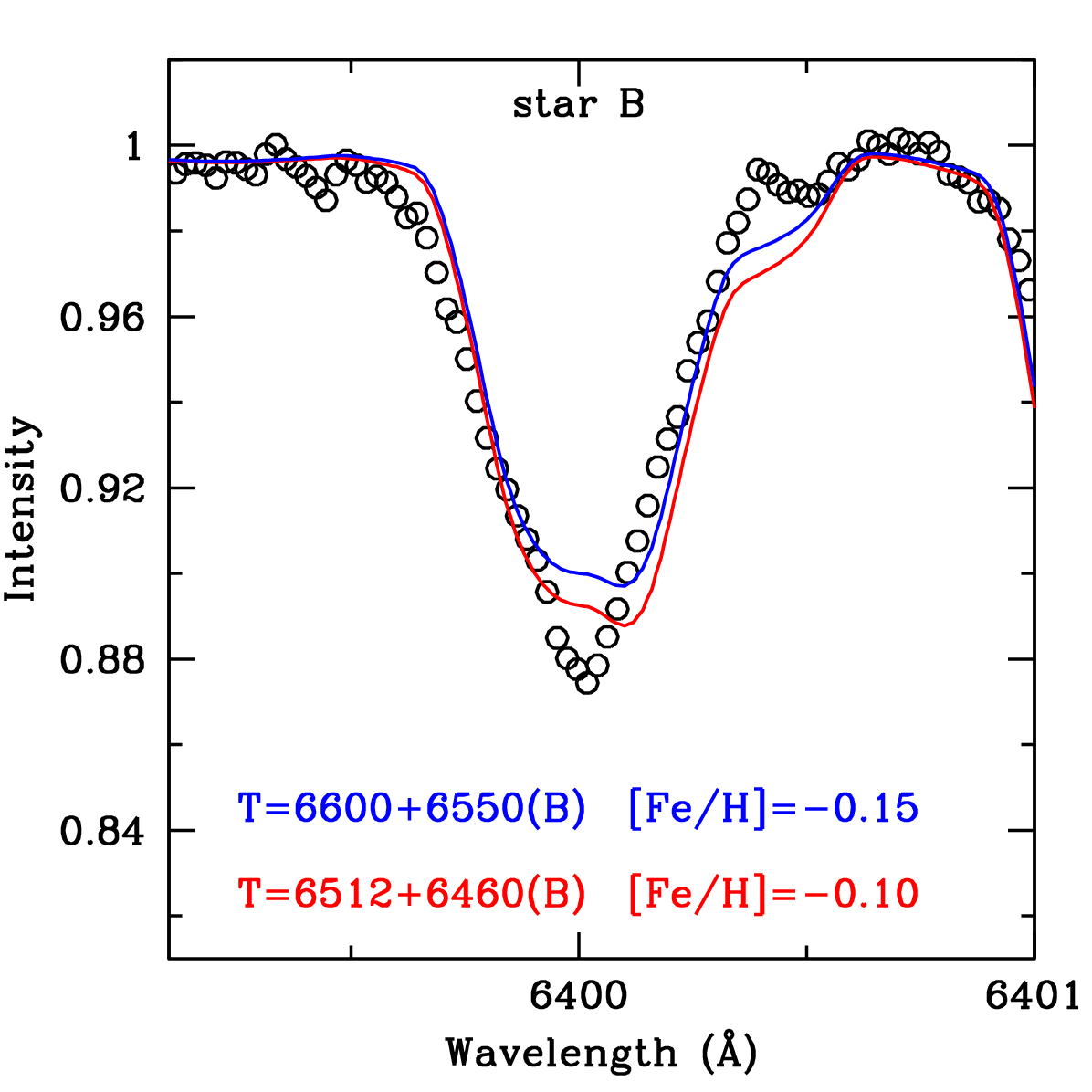}
\includegraphics[width=0.48\linewidth]{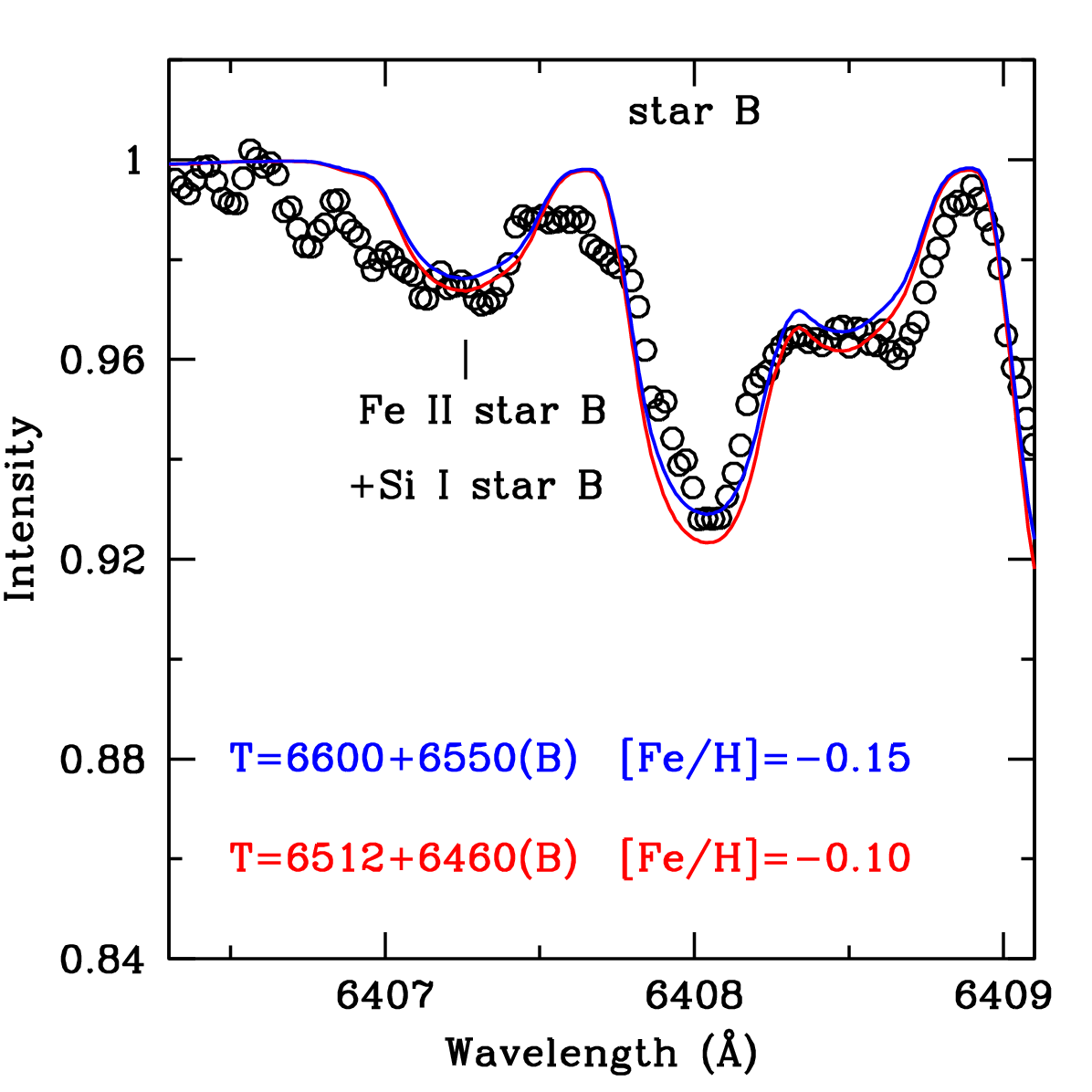}
\includegraphics[width=0.48\linewidth]{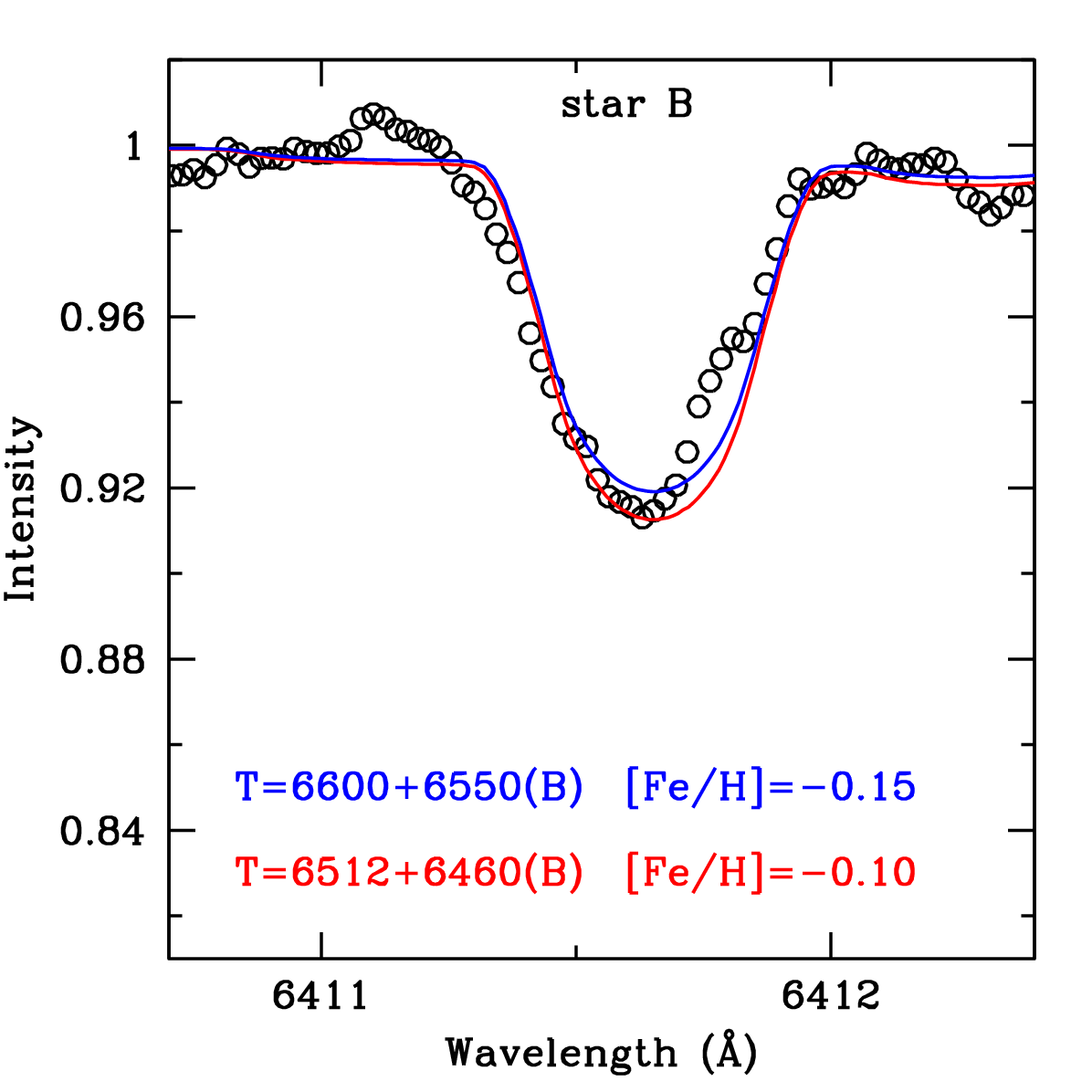}
\caption{Line line profiles of the primary star (dots) compared to model spectra with ($T_{eff}$, $[Fe/H]$)=(6460 K, $-$0.10) (red)
and ($T_{eff}$, $[Fe/H]$)=(6600 K, $-$0.15) (blue).  The model line profiles are broadened to $\vv\sin(i)$=12 km/s.   The $\lambda$6413
line is rather insensitive to $T_{eff}$, while $\lambda$6400.3 shows a relatively strong temperature sensitivity.
This figure shows that models with $[Fe/H]>-$0.1 would not adequately represent the observations.
}
\label{fig_compare_Teff_starB}
\end{figure}

Fig.~\ref{fig_compare_Teff_starA} shows a selection of star A's lines compared to the synthetic line profiles 
from models with either $T_{eff}$=6460\,K or 6600\,K. The higher $T_{eff}$  provides a better match to the observations
and is particularly better for the \ion{Fe}{1} $\lambda$6400 line because of the contribution due to $\lambda$6400.316.
This line contributes to the red wing of the blend. The lower temperature produces synthetic profiles with red wings that 
are clearly too extended. 

The $\lambda$6400.316 line has an excitation potential 0.91 eV, compared to 3.60 eV for the $\lambda$6400.001 
line, making it  more sensitive to temperature. We conclude that, for star A, the best match to the line profile is 
with $T_{eff}$=6650$\pm$50\,K, consistent with the $[Fe/H]=-$0.15 evolutionary track in Fig.~\ref{fig_evolution_tracks}.

%To explore in more detail the temperature dependence of this line profile 
%we generated synthetic spectra for temperatures between 6450K to 6800K with 50K increments for  $[Fe/H]=-$0.15. The
%synthetic spectra were broadened to $\vv\sin(i)$=12 km/s.  The composite spectra were made so as 
%to preserve a 50K difference between the spectra of star A and star B, as required by the temperature ratio which is fixed 
%by the light curve solution. 
%Fig.~\ref{fig_Fe_Teff} shows that a better coincidence between the absorption red wings
%is attained with higher $T_{eff}$.  However, the highest $T_{eff}$ values require a larger $[Fe/H]$ (not shown here) to match 
%the depth of the line core.   

Given the constraint on the effective temperature ratio imposed by the light curve, the above result for 
the star A temperature forces the star B effective temperature to be 6550 K.  Synthetic profiles computed 
with this temperature for Star B are compared
to the observations in Fig.~\ref{fig_compare_Teff_starB}. We find that synthetic spectra with $T_{eff}$ in
the range 6460-6550\,K yield viable fits to the observed profile.  It is not clear whether this larger uncertainty in the
secondary's temperature results from line profiles that are perturbed with respect to predicted line shapes.  What is
clear is that $[Fe/H]>-$0.1 can be excluded for the entire temperature range.  The constraints of the evolutionary
tracks (Fig.~\ref{fig_evolution_tracks}) lead to the conclusion that the higher temperature is to be favored.  

\clearpage
\vfill\eject
\begin{center}
\setlength\LTleft{-2cm}
\begin{longtable}{lccccc}
\caption{Results  \label{table_best_fit}} \\   
\toprule                                 
Parameter          & Literature  &\multicolumn{3}{c}{This paper} &                                                         \\
                   &             & Equiv. Width        & Line Prof.     & Line Prof. &                 \\
\hline
$R_B/R_A$          & 0.98        & 0.98f$^{(a)}$      & 0.98f         & 0.98f      &                      \\
$[Fe/H]_A$         &$-$0.15(.03) &$-$0.166(.07)$^{(b)}$&$-$0.15(.05)  &$-$0.15(.05)&                               \\
$[Fe/H]_B$         &$-$0.12(.03) &$-$0.239(.07)$^{(b)}$& $-$0.15(.05) &$-$0.15(.05)&                    \\
$T_{Aeff}$(K)     & 6512(50)    & 6512(adopted)      &  6500f         & 6600(50)   &                   \\
$T_{Beff}$(K)     & 6460(50)    & 6460(adopted)      &  6450f         & 6550(50)   &                   \\
$\vv\sin(i)_A$ (km/s)& \nodata   & \nodata            &  12.5(.5)     & 12.5(.5)   &                 \\
$\vv\sin(i)_B$ (km/s)& \nodata   & \nodata            &  12.5(.5)     & 12.5(.5)   &                \\
$A(Li)_A$         & 2.67(.1)     & \nodata            & 2.45(.07)     & 2.65(0.07) &                  \\
$A(Li)_B$         & 2.42(.2)     & \nodata            & 2.25(0.1)     & 2.35(.07)  &                   \\
%$log(L_A/L_\odot^N)$& 0.434(.0013)&                  &               & (j)        &                   \\
%$log(L_B/L_\odot^N)$& 0.399(.0013)&                  &               & (j)        &                 \\
\midrule
\bottomrule
\end{longtable}
 \begin{tablenotes}
% \item{} {(aa) $M_\odot^N$, $R_\odot^N$, $L_\odot^N$ are the nominal solar units given by IAU 2015 Resolution
%B3 (A. Pr\~sa et al., AJ, 152, 41, 2016)}
  \item{} {(a) The "f" next to a value indicates that it was a fixed value.}
  \item{} {(b) Values are the average of the results found from the two {\it IRAF/abfind} fits (solar microturbulent 
velocities 1 km/s and 0.8 km/s).}
 \end{tablenotes}
\end{center}

\section{Lithium abundance}

\begin{figure} [!h !t !b]
%/Users/gloriak1/Pilachowski/Spectra/input_compare_obs_out3_Li_T6500zm0.15 input_compare_obs_out3_Li_T6650zm0.15
\includegraphics[width=0.48\linewidth]{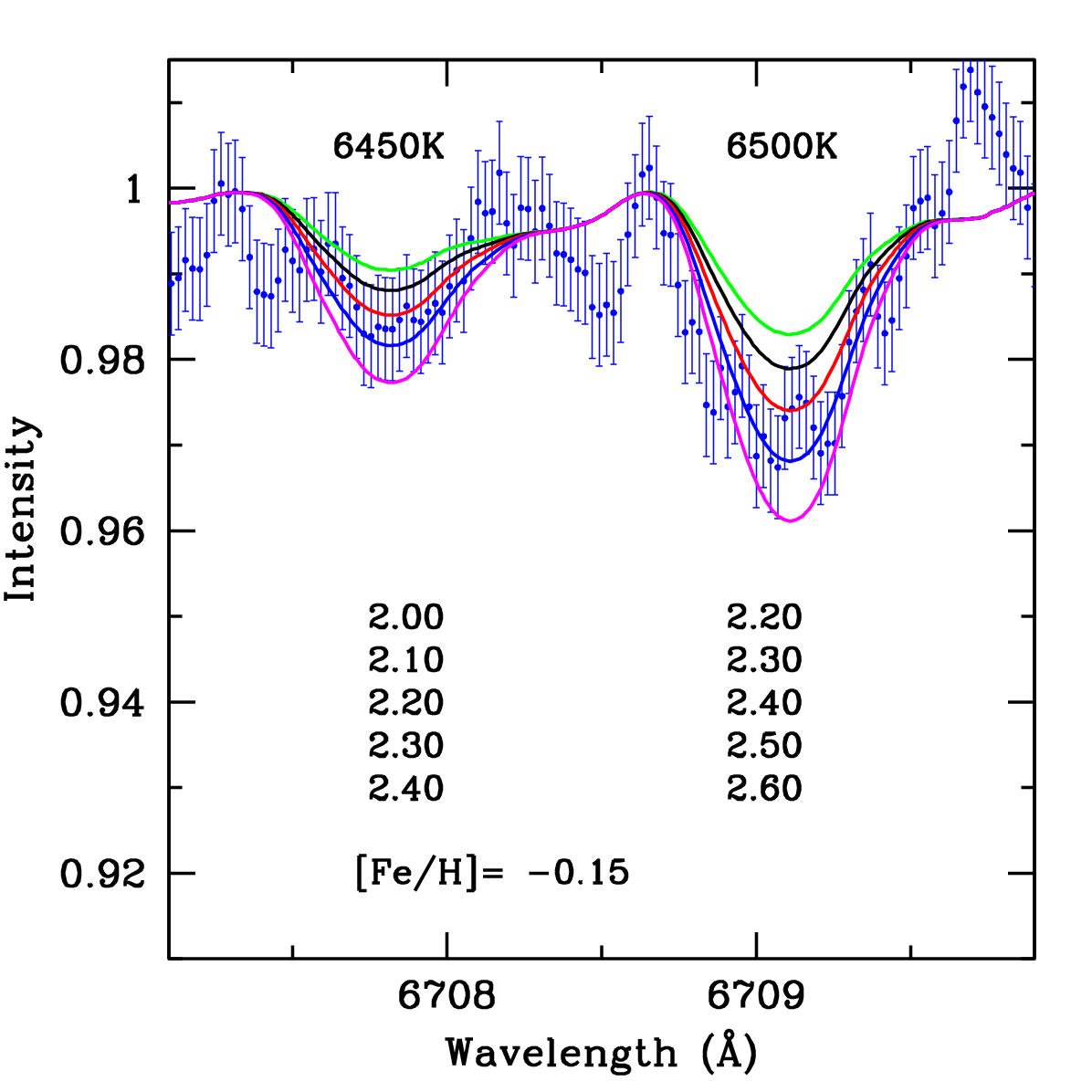}
\includegraphics[width=0.48\linewidth]{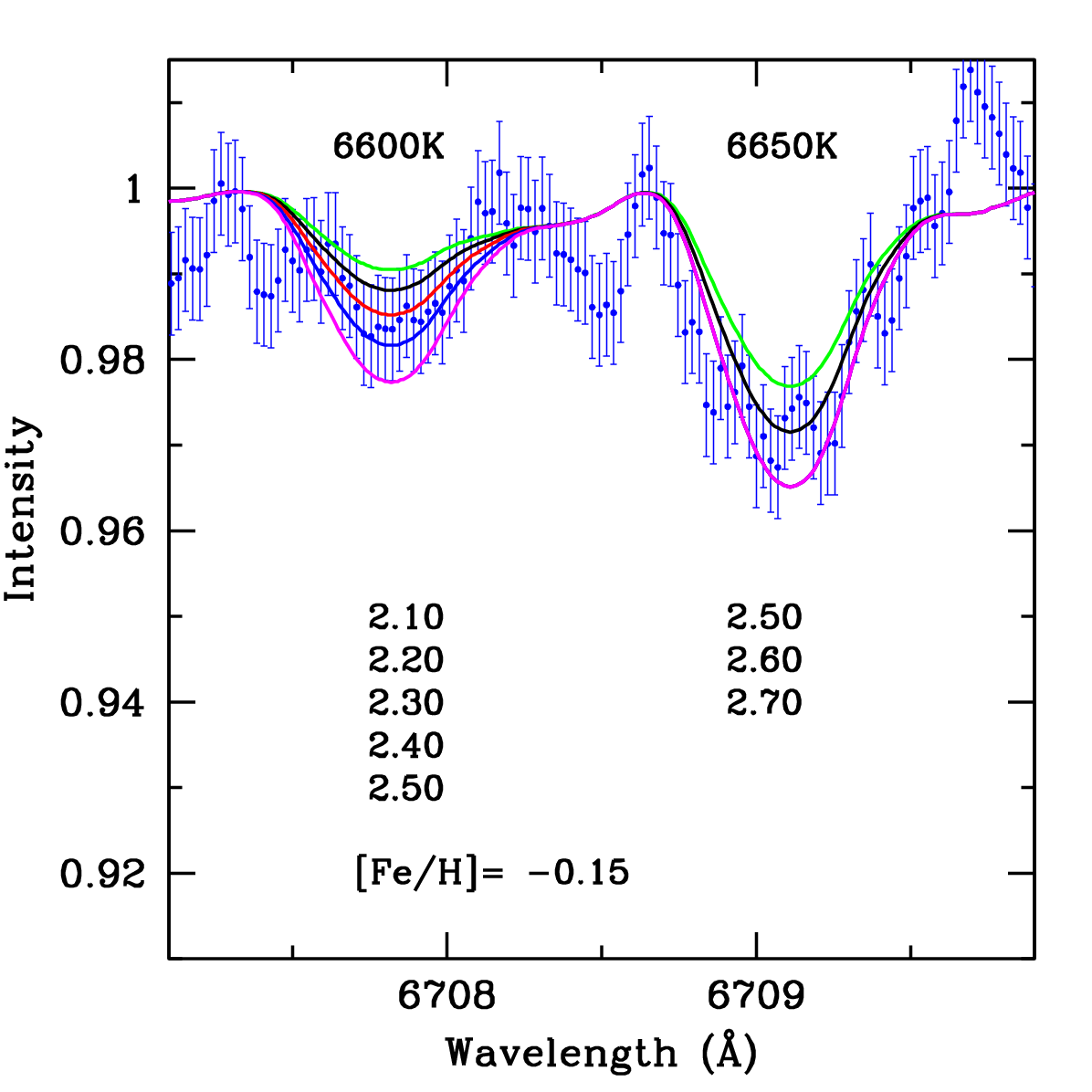}
\caption{Synthetic spectra with different lithium abundances compared to the observed 5-point smoothed 
data (dots). The  model A(Li) values for each star are listed below the correspondng Li line.
The top value (green) corresponds to the curve with the smallest A(Li) and the bottom to the largest.  The error bars
indicate a 0.6\% uncertainty in the 5-point smoothed data. {\bf Left:} Composite model spectra with 
the T08 effective temperatures. {\bf Right:} $T_{eff}$ =(6650 K, 6600 K), for star A and star B, respectively.  
The synthetic spectra are all broadened to $\vv\sin(i)$=12 km/s. 
\label{fig_lithium_abundance}
}
\end{figure}

The Lithium abundance $A(Li)$ was constrained by comparing synthetic spectra constructed with MOOG with the
observed spectrum in the $\lambda\lambda$6704--6710 wavelength range.  We first analyzed $A(Li)$ 
adopting  effective temperatures 6500 K + 6450 K, which approximately correspond to those determined by T08.  
We found $A(Li)$ =2.45$\pm$0.07 (star A) and  2.25$\pm$0.1 (star B), values that are marginally consistent 
with those of Baugh et al. (2013) (2.67$\pm$0.1 and 2.42 $\pm$0.2) and their conclusion that the two components 
seem to have slightly different Li abundances.  The comparison between synthetic spectra and the observations 
for this case is shown in Fig.~\ref{fig_lithium_abundance} (left).

We next analyzed the spectra under the assumption that the stars are hotter, 6650 K+ 6600 K, as implied 
by the $\lambda$6400 line profile. This yielded Lithium abundances $A(Li)$=2.65 $\pm$0.07 
(star A) and 2.35$\pm$0.07 (star B), still consistent with the Baugh et al. (2013) determination.  
The strength of the Li lines does not have a significant dependence on the $[Fe/H]$ value, so our results
are not strongly dependent on the uncertainty in the Iron abundances.  However, it does
depend more strongly on the effective temperature, as shown above.  Our data are unfortunately too
noisy to provide a constraint on the Li abundance to better than 0.05 to 0.1 dex, but both effective
temperature sets do imply a somewhat lower $A(Li)$ in the secondary than in the primary.  
Future observations should
aim for significantly higher S/N spectra to allow for more precise abundance determinations  (as have
been performed for the visual binary components of $\zeta$ Boo A and B by \citet{2022AN....34320036S}).

%%%%%%%%%  Lithium figure %%%%%%%%%%%%%%

\begin{figure} [!h !t !b]
%/Users/gloriak1/Pilacho*/Paper/input_???
\includegraphics[width=0.98\linewidth]{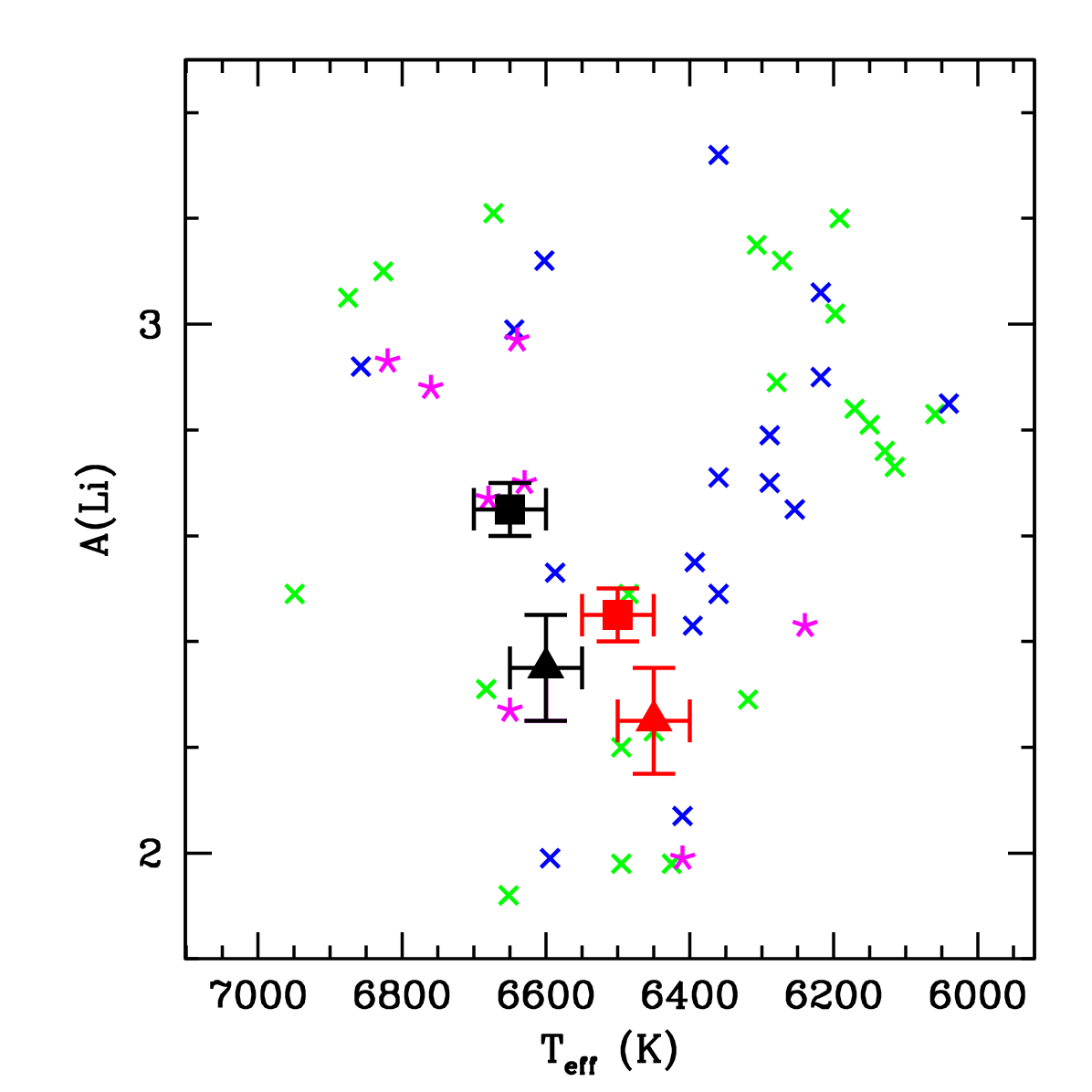}
\caption{Lithium abundance as a function of effective temperature. Crosses indicate data from 
\citet{1995ApJ...446..203B} as follows: Hyades (green, age $\sim$ 625 Myr), Praesepe (blue, age $\sim$680 Myrs),
and NGC752 (magenta, age estimates range from 1.34 to 1.61 Gyr). Our determinations are shown with a
filled-in square for star A and a filled-in triangle for star B. Red/black symbols correspond to the
lower/higher effective temperarutres (see Table 3).
}
\label{fig_Li_vs_Teff}
\end{figure}

\begin{figure} [!h !t !b]
%/Users/gloriak1/Pilacho*/Paper/input_hip10961_RVcurve
\includegraphics[width=0.98\linewidth]{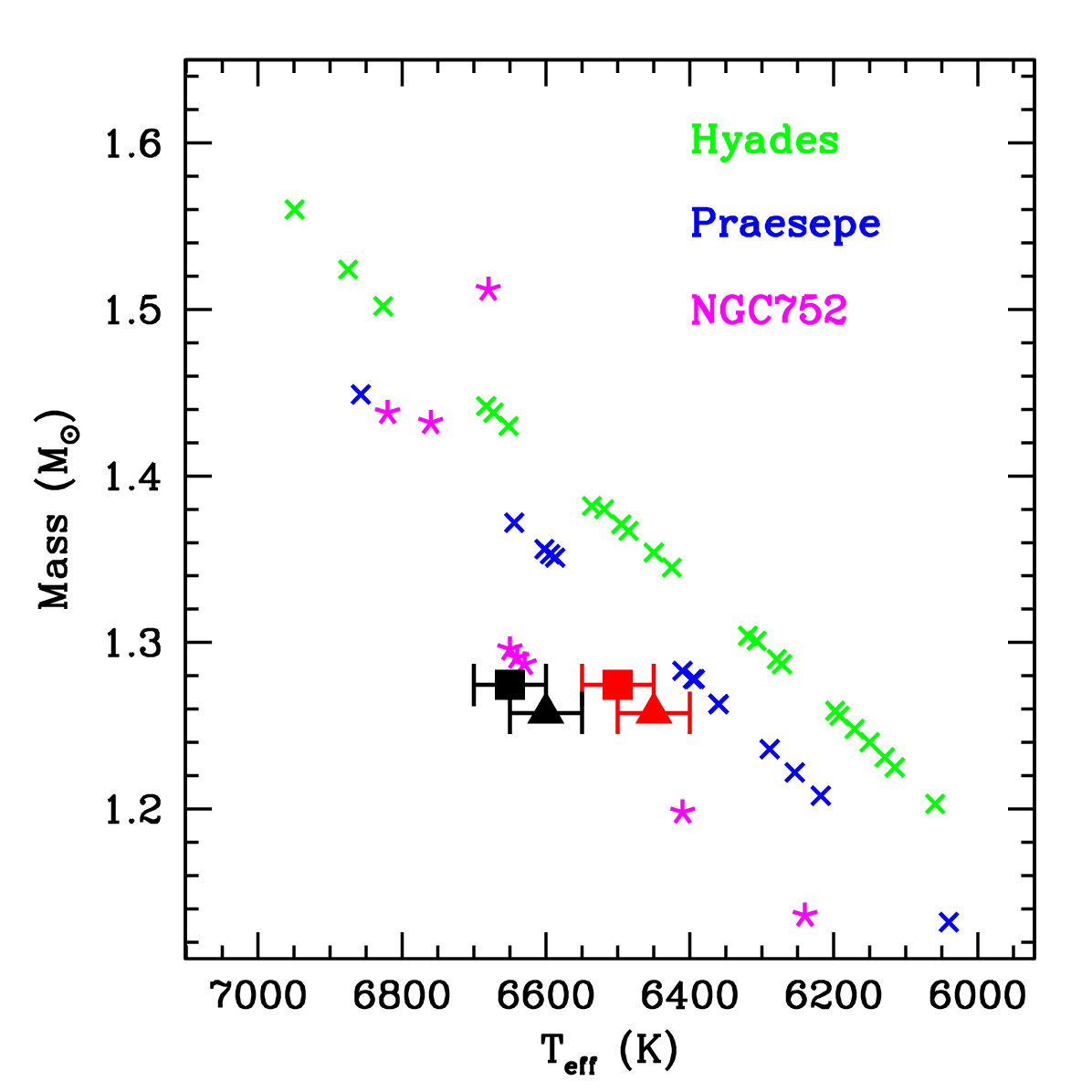}
\caption{The Mass-Effective Temperature data from  \citet{1995ApJ...446..203B} (open squares) for
the same stars as plotted in Fig.~\ref{fig_Li_vs_Teff}.  The filled-in square and triangle as follows: Hyades (green, age $\sim$ 625 Myr), Praesepe (blue, age $\sim$680 Myrs),
and NGC752 (red, age estimates range from 1.34 to 1.61 Gyr). Our star A determinations are shown with a
filled-in square and for starB with a filled-in triangle.
}
\label{fig_Mass_vs_Teff}
\end{figure}

%%%%%%%%%%%%%%%%%% Werner's figure
\begin{figure} [!h !t !b]
% Figs. made by Werner
%\includegraphics[width=0.48\linewidth]{2024oct29_WS_fig1_unscaled.ps}
\includegraphics[width=0.78\linewidth]{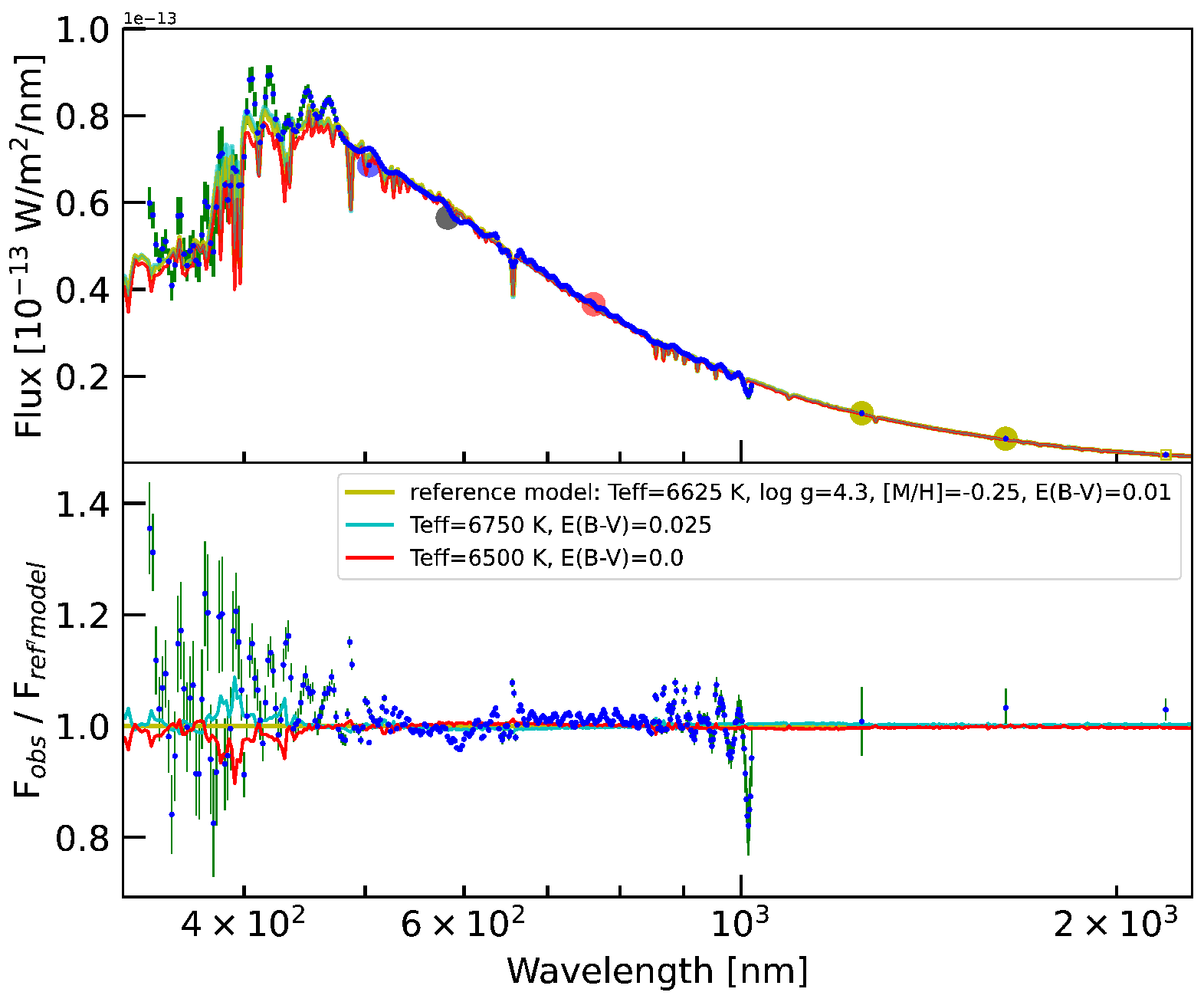}
\caption{Top: Observed spectral energy distribution of V505 Per compared to theoretical SEDs for two stars with 
$T_{eff}$= 6500 K (red), 6625 K (dark green, Reference Model), and 6750 K (light green), and adopting a distance 
$D=62.14 kpc$.  
The observations are the following: large blue, grey and red circles for Gaia Bp, G, and Rp fluxes;
large yellow circles and small blue point for Johnson J, H, and K fluxes;  small blue points with
green error bars for the the Gaia spectral energy distribution (scaled by a factor 0.95 to make it consistent
with the Gaia Rp filter flux, see text).
Bottom: Ratios of the observed SED  models with the reference model in the denominator.  Narrow peaks in the 
ratio result from a mismatch of resolution, because the Kurucz model has a higher resolution than the Gaia XP 
low resolution data.  The mismatch is particularly prominent in the hydrogen-line wavelengths.  Also shown
are the ratios of  $T_{eff}=$6500K and 6750 models to the reference model.  Differences between these 
models and the reference model are only visible in the 400 nm region  where the hotter model has a ratio larger
than unity and the cooler less that unity. In this wavelength region the combination of several strong
narrow absorption lines in the Kurucz reference model and relatively large uncertainties in the observations makes it
difficult to judge which model provides the best fit, although it seems likely that that the reference model
with $T_{eff}=$6625 provides the best fit.  Thus, we conclude that the observed SED is consistent with hotter 
$T_{eff}$ than those determined by T08.            
\label{fig_SED_WS}
}
\label{fig_SED}
\end{figure}

\section{Discussion}

In this paper we analyzed a high resolution spectrum of the eclipsing binary V505 Per.  The objective was
to resolve three inconsistencies related to the chemical abundance of the two components: (a) The first is
the subsolar Fe-abundance reported by Baugh et al. (2013), \citet{2011A&A...530A.138C},  found from spectral  
and photometric analyses  which is inconsistent with the approximately solar abundance that was found by 
Southworth (2021) from isochrone fitting; 
(b) The second is the Li-abundance found by Baugh et al. (2013), which is  factors of 2-5 larger than observed in similar
age and temperature single stars which lie in the Lithium Dip; (c) The third is that the secondary's Li-abundance found
by Baugh et al. (2013) is significantly lower than that of the primary, while both stars are expected to have the same
age and therefore the same abundances.

We first analyzed all the available {\it TESS} photometric data to constrain the parameters that can be
derived from the eclipse light curve analysis (Table 1) which, combined with the published solution to the
radial velocity curves yielded a first set of fundamental parameters.  Because the light curve solution
only yields the sum of the stellar radii ($R_A+R_B$) and not the ratio, we used theoretical evolutionary tracks 
to constrain the value of $R_B/R_A$.  We found that the for a fixed $R_A+R_B$ value as derived from the light
curve, neither the $R_B/R_A$ nor the temperature ratio $T_A/T_B$ have a strong dependence on the metallicity,
with both the solar and subsolar options yielding nearly identical values.  However, the actual temperature
has a strong dependence on metallicity. If the previously published temperatures (T08) are adopted, then the 
system's metallicity is at least solar, as concluded by S21, but the system is significantly older (1.5 Gyr 
instead of $\sim$1 Gyr).  However, if instead the $[Fe/H]$ values are as low as determined by Baugh et al (2013) 
and Casagrande et al., then  $T_{eff}$ must be approximately 
$T_{Aeff}\sim$6670\,K and $T_{Beff}\sim$6620\,K. 

We compared our observed spectrum with stellar atmosphere models with $T_{eff}/K$ in the range [6400, 6800] and
$[Fe/H]$ in the range [-0.4, -0.1] and found that $[Fe/H]$ values larger than $-$0.1 can be excluded.  Specifically,
for the primary/secondary star we found $[Fe/H]$=$-$0.17$\pm$0.07/$-$0.25$\pm$0.07  from \ion{Fe}{1} equivalent widths
(Section 4.1).  Comparing synthetic and observed line profiles, we found $[Fe/H]\sim-$0.13$\pm$0.05 (Figs.~\ref{fig_compare_Teff_starA}
and \ref{fig_compare_Teff_starB}).  Most of the primary star's line profiles are best reproduced with $T_{eff}\sim$6600\,K and
$[Fe/H]=-$0.15 (Fig.~\ref{fig_compare_Teff_starA}).  From the temperature ratio which is fixed by the light curve solution, this
implies that the secondary's effective temperature is $\sim$6550\,K.  However, we find that synthetic spectra with $T_{eff}$ in 
the range 6460-6550\,K yield viable fits to the observed profile.  It is not clear whether the larger uncertainty in the
secondary's derived parameters results from line profiles that seem somewhat perturbed with respect to predicted line shapes.

Having found that temperatures higher than the T08 values are viable,  we then proceded to fit synthetic lithium line profiles
to the observations.  The only well-resolved \ion{Li}{1} lines in our spectrum are those that conform the $\lambda$6707.83 blend.  The
strength of these lines is significantly more dependent on $T_{eff}$ than on $[Fe/H]$. Thus, we modeled the Li lines with the T08
temperatures (for consistency with the results of Baugh et al.) as well as with temperatures that are 150\,K higher.  We found
$A(Li)$ that are consistent with those determined by Baugh et al., within their uncertainties and we also confirmed that the secondary's
lithium abundance is slightly lower than that of the primary. 

\subsection{Higher temperatures eliminate the discrepancy with the Lithium Dip}

A higher $T_{eff}$ places the stars at the hot limit of the Lithium dip  where there is a steep rise in $A(Li)$,
eliminating the discrepancy between their $A(Li)$ and that of cluster stars of a similar age. This is
illustrated in Fig.~\ref{fig_Li_vs_Teff}, where we plot $A(Li)$ {\it vs.} $T_{eff}$ from
\citet{1995ApJ...446..203B} for the  Hyades cluster (age $\sim$ 625 Myr), Praesepe (age $\sim$680 Myrs), and
and NGC752 (age estimates range from 1.34 to 1.61 Gyr).

Higher temperatures are also  more in line with the Mass-$T_{eff}$ relation for NGC 752 than with that
of the younger clusters, as shown in Fig.~\ref{fig_Mass_vs_Teff}, where we plot the same stars as in
Fig.~\ref{fig_Li_vs_Teff} with the data from \citet{1995ApJ...446..203B}.  
%However, if NGC 752 is 1.61 Gyr,  the position of V505 Per using the T08 temperatures is likely better.

Lithium is destroyed by proton capture in stellar interiors at a temperature $\sim$2.5 million
degrees.  In main sequence solar-type stars (M$\sim$1.0$\pm$0.10 M$\odot$), this temperature
is reached below the base of the surface convective zone,  which makes it unlikely for Li-depleted
material to reach the surface unless there is an additional mechanism that transports it to the
convective zone. In the absence of such a mechanism, the surface Li abundance during the main sequence 
is not  predicted to be anomalous.  However, evidence for the existence of such an additional mechanism 
is clearly found, for example, in plots of $A(Li)$ {\it versus} effective temperature T$_{eff}$ in main 
sequence cluster stars with 6300$< T_{eff} < $6900 K \citep{1986ApJ...303..724B, 2002ApJ...565..587B}
in what is called the {\it Li Dip}. The mechanism responsible for the excess mixing during the main sequence 
as thought to be associated with rotation.  In binaries, it has long been suspected that the presence
of a companion can impact the manner in which the mixing process procedes.  This is particularly
true for tidally locked systems in which differential rotation and strong currents should be
suppressed, thus inhibiting strong mixing.  On the other hand, asynchronous binaries may provide a mechanism 
by which enhanced mixing could take place as a consequence of local differential rotation gradients 
\citep{2013A&A...556A.100S, 2021A&A...653A.127K}. 

Our results indicate that
both components in V505 Per rotate subsynchronously.  However, this result is based on a single
orbital phase, near conjunction.  The {\it TESS} light curve displays a clear ellipsoidal effect,
implying that the stars are distorted from a purely spherical shape.  Thus, observations at other
orbital phases are required to determine whether the stars are truly in subsynchronous rotation or
if they undergo line profile variability which, near conjunction, makes the line profiles appear
narrower.

\subsection{Higher temperatures are still consistent with the SED}

A higher effective temperature has an impact on the spectral energy distribution (SED), so the natural question that
arises is whether the  SED  of a  $T_{eff}$=6650 K + 6550 K system is consistent with the observations.
\footnote{There can be differences of up to 200 K in the effective temperatures as determined photometrically, 
depending on the calibrations that are used, and this has an impact on the $[Fe/H]$ values, see for example 
Balachandran (1995). NOTE: this author (and others?) plot A(Li) vs. $T_{ZAM}$.  This means that for the older
clusters they have to assume an evolutionary model predicting the T$_{ZAMS}$ given the current Teff and the
age of the system.}

We now show that higher $T_{eff}$ values are consistent with the observed SED.  In Fig.~\ref{fig_SED_WS} we plot the 
observed absolute flux in the different wavelength bands for which it is available in the literature.  Specifically, 
we use the  Bp, G, and Rp fluxes, absolute flux calibrated as given in Vizier \citep{vizier2000}, referring to the GAIA 
DR3 data release \citep{Riello2021}, to the 2MASS sky survey \citep{2MASS2006} for the Johnson filter $J, H, K$ 
measurements (nominal effective wavelengths 1250\,nm, 1630\,nm, and 2190\,nm respectively), and the Gaia XP 
spectrum \citep{XPspectra2023}. The latter is approximately 5\,\% brighter than the other Gaia fluxes,
so we scaled the  XP spectrum by 0.95 to bring it into agreement with the other Gaia fluxes. The absolute fluxes 
are compared to composite model SEDs for three effective temperatures assuming stellar radii as determined by 
S21 (1.294 $R_\odot$ and 1.264 $R_\odot$) and a distance $D=62.14\pm 0.12$\,pc, which  results from a Gaia DR3 
parallax of $\pi=16.068\pm 0.02$\,mas and a zero point correction of $Z5=-0.024$\,mas according to the recipe 
given by \citep{Lindegren_etal2021}.\footnote{S21 determined a distance $D$=61.19$\pm$0.62 pc to the system, 
while the {\it Gaia} (re-interpreted) EDR3 parallaxes give 62.03$\pm$0.10pc (C. A. L. Bailer-Jones et al., 
AJ, 161, 147, 2021).  }
%wks 14.11.24 Remark: these distances agree within the combined given precision
% the distance of Bailer-Jones does probably not account for the zero point correction (I have not checked)

% wks moved: and reddening  E(B-V)=0.035.
The models were interpolated for $\log(g)$=4.3, $[M/H]$=$-0.25$ and $T_{eff}$= 6500\,K, 6625\,K, and 6750\,K 
from the Atlas9 Castelli-Kurucz grids \citep{2003IAUS..210P.A20C}.
\footnote{https://wwwuser.oats.inaf.it/fiorella.castelli/grids/gridp00k2odfnew/fp00k2tab.html}.

%       editor = {{Piskunov}, N. and  {Weiss}, W. W. and {Gray}, D. F.},

%wks 14.11.2024 added
For each of the effective temperatures, the SEDs are reddened, respectively, by $E(B-V)=0$, %($T_{eff}$= 6500\,K),
$=0.02$, and $=0.03$ %($T_{eff}$= 6750\,K).
It can be seen that there is no significant difference between these models and the observed energy distribution.

However, the obvious difference between the models and the absolute fluxes is that the models have about a 3\,\%
too small flux in the infrared. The dominant uncertainty at these wavelengths are the Johnson filter measurements, 
which for the $K$ fileter is $\pm 2$\,\%. Thus, the difference is statistically not significant but it seems to be real 
because it is for both, the $H$ and $K$ filters about the same ratio. Any absolute difference at optical wavelengths 
can be compensated by the reddening.

Future observations should
aim for significantly higher S/N spectra to allow for more precise abundance determinations, as have
been performed for the visual binary components of $\zeta$ Boo A and B by Strassmeier \& Steffen (2022, AN).

\section{Conclusions}

Adopting the $T_{eff}$ values from T08, Baugh et al. (2013) determined  Lithium abundances that are significantly
higher than the Lithium abundance in similar-age cluster stars of the same temperatures.  They also found that star A
and star B  have marginally different $A(Li)$.  We repeated this analysis with our R=100,000, S/N$\sim$100 echelle
spectrum and confirm these results.  However, our analysis of the \ion{Fe}{1} line profiles  suggests
that $T_{eff}$ could actually be larger by at least 150K than the values determined by T08. Allowing for hotter 
$T_{eff}$, we find that both stars lie near the hot edge of the Lithium Dip where such abundances aren't uncommon.  
Higher $T_{eff}$ values also help relieve the tension between the model line profiles of the $\lambda$6400.35 
absorption line since for lower temperatures the red wing of the line profile is significantly stronger than observed.
In addition,  the combined constraints of the eclipse light curve and the evolutionary tracks require higher
$T_{eff}$ values in order to satisfy the results obtained by us and other authors that  $[Fe/H]$  is subsolar.
The higher effective temperatures suggested by our analysis would still be consistent with the observed spectral 
energy distribution of the V505 Per system, although marginally larger K-band fluxes are observed than predicted
by the models.  If this difference is real, V505 Per may contain an as yet undetected, low mass third object.

Further high S/N and spectral resolution observations are needed to allow for more precise abundance determinations,
as well as to check for orbital phase-dependent variations that could affect the line strengths and profiles, particularly
because we find the stars to be in subsynchronous rotation and hence could be undergoing tidally driven perturbations
which are orbital-phase dependent.

\acknowledgements

Support from UNAM DGAPA PAPIIT program IN105723 and from the Indiana University Institute for Advanced Study 
are gratefully acknowledged. GK and AM thank Andrew Tkachenko for useful comments during the early phases of this
investigation.

%\bibliographystyle{rmaa}
%Gustafsson G., Edvardsson et al., 2008, A\&A, 486, 951.  https://marcs.astro.uu.se

\bibliography{references_V505Per_2025jan10.bib}

\appendix

\end{document}